\documentstyle[12pt]{article}

\newcommand{\nc}{\newcommand}
\newcommand{\rnc}{\renewcommand}

\makeatletter
\def\section{\@startsection {section}{1}{\z@}{-3.5ex plus -1ex minus
 -.2ex}{2.3ex plus .2ex}{\large\bf}}
\def\subsection{\@startsection{subsection}{2}{\z@}{-3.25ex plus -1ex minus
 -.2ex}{1.5ex plus .2ex}{\normalsize\bf}}
\makeatother

\makeatletter
\@addtoreset{equation}{section}
\rnc{\theequation}{\thesection.\arabic{equation}}
\makeatother

\nc{\ignore}[1]{}

\nc{\be}{\begin{equation}}
\nc{\ee}{\end{equation}}
\nc{\bea}{\begin{eqnarray}}
\nc{\eea}{\end{eqnarray}}

\rnc{\a}{\alpha}
\rnc{\d}{\delta}
\nc{\ep}{\epsilon}
\nc{\e}{\eta}
\nc{\eb}{\bar{\eta}}
\nc{\f}{\phi}
\nc{\fb}{\bar{\phi}}
\nc{\vf}{\varphi}
\nc{\p}{\psi}
\rnc{\pb}{\bar{\psi}}
\rnc{\c}{\chi}
\nc{\cb}{\bar{\c}}
\nc{\la}{\lambda}
\nc{\m}{\mu}
\nc{\n}{\nu}
\rnc{\o}{\omega}
\rnc{\t}{\theta}
\nc{\tb}{\bar{\theta}}
\nc{\eps}{\epsilon}
\nc{\Om}{\Omega}
\nc{\ga}{\gamma}

\rnc{\S}{\Sigma_{g}}
\nc{\Sa}{\S\times\{0\}}
\nc{\Sb}{\S\times\{1\}}
\nc{\SI}{\S\times I}
\nc{\SS}{\S\times S^{1}}
\nc{\Sg}{\S}
\nc{\M}{{\cal M}}
\nc{\MF}{\M_{{\cal F}}}

\nc{\trac}[2]{{\textstyle\frac{#1}{#2}}}

\nc{\ex}[1]{{\rm e}^{\,\textstyle#1}}

\nc{\mat}[4]{\left(\begin{array}{cc}#1&#2\\#3&#4\end{array}\right)}

\def\Tr{\mathop{\rm Tr}\nolimits}
\def\tr{\Tr}
\nc{\ra}{\rightarrow}
\nc{\ot}{\otimes}
\rnc{\ss}{\subset}
\nc{\ad}{{\rm ad}}
\nc{\Ad}{{\rm Ad}}
\nc{\nul}{\noindent\underline}

\rnc{\lg}{{\bf g}}
\nc{\lt}{{\bf t}}
\nc{\lk}{{\bf k}}
\nc{\lf}{{\bf f}}
\nc{\lh}{{\bf h}}
\nc{\bft}{{\bf t}}
\nc{\bfk}{{\bf k}}
\nc{\bfg}{{\bf g}}
\nc{\del}{\partial}
\nc{\dbar}{\bar{\del}}
\nc{\dz}{\del_{z}}
\nc{\zb}{\bar{z}}
\nc{\dzb}{\del_{\bar{z}}}
\nc{\az}{A_{z}}
\nc{\azb}{A_{\bar{z}}}
\nc{\bz}{B_{z}}
\nc{\bzb}{B_{\bar{z}}}
\nc{\ba}{{\bf A}}
\nc{\bb}{{\bf B}}
\nc{\g}{g^{-1}}
\nc{\dw}{\Delta_{W}}
\nc{\Det}{{\rm Det}\,} 
\nc{\ddw}{\det\dw}
\nc{\Ddw}{\Det\dw}
\nc{\bG}{{\bf G}}
\nc{\bT}{{\bf T}}
\nc{\bF}{{\bf F}}
\nc{\bH}{{\bf H}}
\nc{\unA}{\underline{A}}
\nc{\C}{{\cal A}/{\cal G}}
\nc{\A}[1]{{\cal A}^{#1}/{\cal G}^{#1}}
\nc{\dx}{\dot{x}}
\rnc{\O}[2]{\Omega^{#1}({#2},\lg)}
\nc{\wif}{Weyl integral formula}
\nc{\CS}{Chern-Simons}

\def\sAA{{\rm A\kern-0.85em A}} 
\def\tAA{{\mathchoice
  {\sAA}
  {\sAA}
  {\rm A\kern-0.60em A}
  {\rm A\kern-0.50em A} }}
\def\sBB{{\rm I\kern-.17em{}B}}
\def\BB{{\mathchoice
  {\sBB}
  {\sBB}
  {\rm I\kern-.13em{}B}
  {\rm I\kern-.13em{}B} }}
\def\sCC{{\kern 0.27em\vrule height1.45ex width0.03em depth0em
          \kern-0.30em\rm C}}
\def\CC{{\mathchoice
  {\sCC}
  {\sCC}
  {\kern 0.225em \vrule height1.05ex width0.025em depth0em \kern-0.25em \rm C}
  {\kern 0.180em \vrule height0.78ex width0.02em depth0em \kern-0.2em \rm C}
        }}
\def\tCC{{\ooalign{C\crcr\kern0.27em\vrule height1.45ex width0.03em
depth0em\crcr}}}
\def\sDD{{\rm I\kern-.16em{}D}}
\def\DD{{\mathchoice
  {\sDD}
  {\sDD}
  {\rm I\kern-.13em{}D}
  {\rm I\kern-.13em{}D} }}
\def\sEE{{\rm I\kern-.17em{}E}}
\def\EE{{\mathchoice
  {\sEE}
  {\sEE}
  {\rm I\kern-.13em{}E}
  {\rm I\kern-.13em{}E} }}
\def\sFF{{\rm I\kern-.16em{}F}}
\def\FF{{\mathchoice
  {\sFF}
  {\sFF}
  {\rm I\kern-.13em{}F}
  {\rm I\kern-.13em{}F} }}
\def\sGG{{\kern 0.27em \vrule height1.45ex width0.03em depth0em
          \kern-0.30em \rm G}}
\def\GG{{\mathchoice
  {\sGG}
  {\sGG}
  {\kern 0.225em \vrule height1.05ex width0.025em depth0em \kern-0.25em \rm G}
  {\kern 0.180em \vrule height0.78ex width0.020em depth0em \kern-0.20em \rm G}
        }}
\def\sHH{{\rm I\kern-.16em{}H}}
\def\HH{{\mathchoice
  {\sHH}
  {\sHH}
  {\rm I\kern-.13em{}H}
  {\rm I\kern-.13em{}H} }}
\def\sII{{\rm I\kern-.16em{}I}}
\def\II{{\mathchoice
  {\sII}
  {\sII}
  {\rm I\kern-.12em{}I}
  {\rm I\kern-.10em{}I} }}
\def\sJJ{{\kern0.17em\vrule height1.5ex width 0.03em depth0em
          \kern-.20em\rm J}}
\def\JJ{{\mathchoice
  {\sJJ}
  {\sJJ}
  {\kern0.150em\vrule height1.05ex width 0.025em depth0em\kern-.175em\rm J}
  {\kern0.135em\vrule height0.78ex width 0.020em depth0em\kern-.155em\rm J} }}
\def\sKK{{\rm I\kern-.16em{}K}}
\def\KK{{\mathchoice
  {\sKK}
  {\sKK}
  {\rm I\kern-.12em{}K}
  {\rm I\kern-.10em{}K} }}
\def\sLL{{\rm I\kern-.16em{}L}}
\def\LL{{\mathchoice
  {\sLL}
  {\sLL}
  {\rm I\kern-.12em{}L}
  {\rm I\kern-.10em{}L} }}
\def\sMM{{\rm I\kern-.16em{}M}}
\def\MM{{\mathchoice
  {\sMM}
  {\sMM}
  {\rm I\kern-.12em{}M}
  {\rm I\kern-.10em{}M} }}
\def\sNN{{\rm I\kern-.16em{}N}}
\def\NN{{\mathchoice
  {\sNN}
  {\sNN}
  {\rm I\kern-.12em{}N}
  {\rm I\kern-.10em{}N} }}
\def\sOO{{\kern 0.27em \vrule height1.50ex width0.03em depth0em
					\kern-0.30em \rm O}}
\def\OO{{\mathchoice
  {\sOO}
  {\sOO}
  {\kern 0.225em \vrule height1.05ex width0.025em depth0em \kern-0.25em \rm O}
  {\kern 0.180em \vrule height0.78ex width0.020em depth0em \kern-0.20em \rm O}
        }}
\def\sPP{{\rm I\kern-.16em{}P}}
\def\PP{{\mathchoice
  {\sPP}
  {\sPP}
  {\rm I\kern-.12em{}P}
  {\rm I\kern-.10em{}P} }}
\def\sQQ{{\kern 0.27em \vrule height1.45ex width0.03em depth0em
          \kern-0.30em \rm Q}}
\def\QQ{{\mathchoice
	{\sQQ}
	{\sQQ}
  {\kern 0.225em \vrule height1.05ex width0.025em depth0em \kern-0.25em \rm Q}
  {\kern 0.180em \vrule height0.78ex width0.020em depth0em \kern-0.20em \rm Q}
        }}
\def\sRR{{\rm I\kern-0.16em{}R}}
\def\RR{{\mathchoice
  {\sRR}
  {\sRR}
  {\rm I\kern-0.12em{}R}
  {\rm I\kern-0.10em{}R} }}
\def\sSS{{\rm S\kern-.45em{}S}}
\def\sTT{{\rm T\kern-.60em{}T}}
\def\TT{{\mathchoice
  {\sTT}
  {\sTT}
  {\rm T\kern-.45em{}T}
  {\rm T\kern-.38em{}T} }}
\def\sUU{{\rm U\kern-.60em{}U}}
\def\UU{{\mathchoice
  {\sUU}
  {\sUU}
  {\rm U\kern-.46em{}U}
  {\rm U\kern-.38em{}U} }}
\def\sVV{{\rm V\kern-.62em{}V}}
\def\VV{{\mathchoice
  {\sVV}
  {\sVV}
  {\rm V\kern-.46em{}V}
  {\rm V\kern-.38em{}V} }}
\def\sWW{{\rm W\kern-.92em{}W}}
\def\WW{{\mathchoice
  {\sWW}
  {\sWW}
  {\rm W\kern-.80em{}W}
  {\rm W\kern-.67em{}W} }}
\def\sXX{{\rm X\kern-.58em{}X}}
\def\XX{{\mathchoice
  {\sXX}
  {\sXX}
  {\rm X\kern-.45em{}X}
  {\rm X\kern-.38em{}X} }}
\def\sYY{{\rm Y\kern-.58em{}Y}}
\def\YY{{\mathchoice
  {\sYY}
  {\sYY}
  {\rm Y\kern-.45em{}Y}
  {\rm Y\kern-.40em{}Y} }}
\def\sZZ{{\rm Z\kern-0.32em{}Z}}
\def\ZZ{{\mathchoice
  {\sZZ}
  {\sZZ}
  {\rm Z\kern-0.30em{}Z}
  {\rm Z\kern-0.25em{}Z} }}

\begin{document}
\global\parskip=4pt


\begin{titlepage}
\newlength{\titlehead}
\settowidth{\titlehead}{NIKHEF-H/91}
\begin{flushright}
\parbox{\titlehead}{
\begin{flushleft}
IC/93/356\\
October 1993\\
hep-th/9310144 \\
\end{flushleft}
}
\end{flushright}
\begin{center}
\vskip 1.5in
{\LARGE\bf Lectures on 2d Gauge Theories}\footnote{Presented at the 1993
Trieste Summer School in High Energy Physics and Cosmology, 14 June - 30
July 1993.} \\
\vskip .4in
{\Large\bf Topological Aspects and}\\
\vskip .25in
{\Large\bf Path Integral Techniques}
\vskip 1.5in
{\bf Matthias Blau}\footnote{e-mail: blau@ictp.trieste.it}
and {\bf George Thompson}\footnote{e-mail: thompson@ictp.trieste.it}\\
\vskip .10in
ICTP\\
P.O. Box 586\\
I-34014 Trieste\\
Italy
\end{center}
%
\end{titlepage}


\tableofcontents
\setcounter{footnote}{0}

\section{Introduction}

In these lectures we will deal with two classes of two-dimensional field
theories which are not obviously topological (in the more traditional sense of
the word) but which nevertheless exhibit an intriguing equivalence with
certain topological theories. These classes are two-dimensional Yang-Mills
theory and the so-called $G/G$ gauged Wess-Zumino-Witten model. The aim of
these lectures will be
\begin{enumerate}
\item to exhibit and extract the topological information that is contained
in these theories, and
\item to present a technique which allows one to calculate directly their
partition function and topological correlation functions on arbitrary
closed surfaces.
\end{enumerate}
As
the claim that two-dimensional Yang-Mills theory
is in some sense topological may seem
somewhat bizarre, let us make more precise what we mean by `topological'
in the present context. First of all, what both Yang-Mills theory in 2d
and the $G/G$ model have in common is that they have no field theoretic
degrees of freedom (i.e.~the spectra of these theories contain no particles
but only `global' excitations of the fields). Moreover, in both of those
theories one can find a fundamental class of correlation functions of local
scalar operators which are independent of the points at which these operators
are evaluated.\footnote{In addition, there are correlators
which display an analogous higher homological invariance.}
In particular this means that these correlators will depend only on the
genus of the surface (and on the area of the surface in the case of
Yang-Mills theory). It is
this invisibility of a background structure which is the hallmark of
a topological theory. In the case at hand these correlation
functions turn out to be related to the topology of the moduli space of
flat connections on a 2d surface $\S$ of genus $g$.

As we will see that in a
precise sense Yang-Mills theory (or rather its topological
zero-coupling limit, 2d BF theory) can be regarded as a tangent space
approximation of the $G/G$ model, being based on the Lie algebra of
$\bG$ instead of on $\bG$ itself, all the topological information
could be obtained directly from the $G/G$ model in a certain
limit. It is instructive, however, to deal with Yang-Mills theory
seperately first, so as to introduce and test the techniques we use in
these lectures in a well-understood example.

As a preliminary step towards extracting the topological information from
these theories, one requires an explicit expression for the partition function
on $\S$ (since, as we shall see, it can be regarded as a generating functional
for the topological correlators). For Yang-Mills theory in 2d such an
expression
has been obtained previously by various methods \cite{ewym,btym,rusakov}, all
requiring, however, in one way or another lattice gauge theory or cut-and-paste
techniques involving calculations on manifolds with boundary as an
intermediate step. This technique does not extend directly to the $G/G$ model
(because the definition of the Wess-Zumino-Witten model on a manifold with
boundary is not straightforward), and it is thus desirable to develop other
techniques which do allow one to deal with closed surfaces directly. The method
we want to advocate here, which can perhaps loosely be referred to as
Abelianization, can be applied to both Yang-Mills theory (where it reproduces
in a straightforward manner the known result - with an interesting twist)
and the $G/G$ model. It is based on a particular gauge choice which permits
one to seperate the Abelian from the non-Abelian components of the gauge
(and other) fields and to eliminate the latter, leaving one with a quadratic
Abelian theory instead of the original non-linear non-Abelian theory. This
purely Gaussian theory is easily solved.

In the case of Yang-Mills theory, this way of proceding is closely related
to the work of Witten \cite{ew2d} who gave another derivation of the
Yang-Mills partition function in terms of a non-Abelian localization theorem,
expressing the result as the sum over contributions from the classical
solutions of the Yang-Mills equations. In fact, our derivation can be regarded
as an alternative derivation of Witten's non-Abelian localization theorem
in this particular case by combining Abelianization with Abelian localization
(a technique that has been used recently in \cite{jk}
to give a rigorous proof of this theorem in a finite dimensional setting).

As this procedure of Abelianization
is a direct path integral analogue of what is known as the Weyl integral
formula in Lie group theory (and its Lie algebra counterpart), after having
introduced the models themselves we spend some time explaining these ideas
in the classical finite dimensional context. For reference purposes
and to establish our notation we have collected the relevant facts from Lie
algebra theory in Appendix A. As to the more detailed structure
of these lecture notes we refer to the table of contents. By and large,
these lecture notes are based on the papers \cite{ewym,ew2d} by Witten
(section 2) and on \cite{btver} (section 3).

We should also mention some of the things we do not cover
(but perhaps should have covered) in these notes. First of all,
these lectures are not meant to be an introduction to Topological
Field Theory per se - for these see e.g.~the review \cite{pr} or
previous procedings of this Summer School. We also say very little
about the relation between the $G/G$ model and Chern-Simons theory -
this has been discussed in \cite{btver}.
Moreover, although we
will be mentioning things like conformal blocks and fusion rules,
we have attempted to avoid the use of conformal field theory, relying
on more elementary gauge theory techniques instead. Indeed, one of
the motivations for pursuing this approach was the possibility to
derive the Verlinde formula, a deep result in conformal field theory,
from Chern-Simons theory without knowing the first thing about conformal
field theory. Finally, we should mention that the technique of
Abelianization can also be applied to higher-dimensional (topological)
field theories like Chern-Simons and BF theories, at least on manifolds
of the form $N\times S^{1}$.

\section{Yang-Mills and Topological Gauge Theories in Two Dimensions}

In this section we will discuss in some detail the evaluation of the
Yang-Mills path integral and its relation to a topological field theory.
The main references for the topological
interpretation of the results are \cite{ewym,ew2d},
while some of the techniques are based on those used in \cite{btver}. For more
background on the topological theories in question see for example
\cite{pr,btbf2,gtlec}.

\subsection{Yang-Mills Theory in Two Dimensions}

The Yang-Mills action in two dimensions is
\be
S = \frac{1}{8\pi^{2}\ep}\int_{\S } \Tr F_{A}*F_{A} \, ,
\ee
with the trace taken in the fundamental representation (for $SU(n)$) and
the matrix connection (gauge field) $A$ is anti-Hermitian.
The path integral that we would like to compute is then,
\bea
Z_{\Sg}( \ep) &=& \int DA \exp{ \left(\frac{1}{8 \pi^{2}\ep}\int_{\S } \Tr
F_{A}*F_{A}  \right)}
\nonumber \\
& =& \int DA D\f \exp{ \left( \frac{1}{4\pi ^{2}} \int_{\S} \Tr i\f F_{A}
+ \frac{\ep}{8 \pi^{2}} \int_{\S} d\m \Tr \f^{2} \right)} \, , \label{pi}
\eea
where the second line is seen to imply the first on performing the
Gaussian integral over $\f$. Here $\f$ is taken to be an anti-Hermitian
matrix\footnote{The reader should be warned that when we come to the ${\bf
G}/{\bf G}$ models $\f$ will denote a Hermitian, though compact, field.}
and $\ep$ plays the role of the coupling
constant. All the other factors have been chosen so as to normalise
things in accord with fixed point theorems.

The gauge invariance of the action in the form that appears in
(\ref{pi}) is
\be
A' = g^{-1}Ag + g^{-1}dg \, , \; \; \; \; \; \;\; \f' = g^{-1}\f g \, .
\ee

\noindent \underline{Metric Dependence}

The metric enters in (\ref{pi}) only in
terms of the measure (or area element) $d\m$. If we scale the coupling
constant $\ep$ by $\lambda$, that is $\ep \ra \lambda \ep$ then this can
be compensated by scaling the metric $g_{\m \n}$ by $\lambda^{-1}$, for
then $d\m \ra \lambda^{-1} d\m$. This means, in turn, that the path
integral will only be a function of the product $\ep A_{\S}$, where
$A_{\S}$ is the area of the surface. As this is the case we may as
well work with a metric of unit area and we do so henceforth. All the
`metric' dependence is then to reside in $\ep$.  We also
introduce a symplectic two-form $\o$ with unit area, for later use,
which allows us to write
\be
\int_{\S} d\m \Tr \f^{2} = \int_{\S} \, \o \Tr \f^{2} \, .
\ee

Yang-Mills theory in two dimensions is then almost a topological theory.
The action (\ref{pi}) is invariant under all diffeomorphisms (general
co-ordinate transformations) which leave the area fixed. It becomes a
topological theory when we set $\ep =0$ or, equivalently, when the
surface is allowed to degenerate to zero area.

\noindent \underline{Observables}

A standard set of physical observables in Yang-Mills theory are, in any
dimension, Wilson loops. These are, of course interesting observables
for the two dimensional theory and have been analysed to a great extent
in e.g.~\cite{btym,gross,loops}.
The methods that we shall employ to evaluate the partition
function are, however, more naturally suited to dealing with another set
of observables. These are any gauge invariant polynomials of the field
$\f$. For example, products of $\Tr \f^{2}(x_{i})$ at various points
$x_{i}$ are the type of observables that we have in mind. More generally
one may consider any homogeneous invariant polynomial $P_{n}(x)$, of
degree $n$, on the Lie algebra $\lg$ as a basic observable.

The almost topological nature of Yang-Mills theory in two dimensions
becomes even more apparent when we study the dependence of the
observables on their positions on $\Sg$. One may use a Schwinger-Dyson
equation (variation with respect to the gauge field $A$)
to show that expectation values of products of the $P_{n}$,
\be
< \, \prod_{i=1}^{m} \, P_{n_{i}}(x_{i}) \,>_{\ep} \, , \label{pob}
\ee
do not depend on the point at which one is evaluating them. Our
conventions are that the expectation value of an operator ${\cal O}$ in
Yang-Mills theory is denoted by
\be
<{\cal O}>_{\ep} \equiv \int_{\Phi} \, {\cal O} \, \exp{ \left(
\frac{1}{4\pi ^{2}} \int_{\S} \Tr i\f F_{A}
+ \frac{\ep}{8 \pi^{2}} \int_{\S} d\m \Tr \f^{2} \right)} \,
\ee
where the symbol $\int_{\Phi}$ stands for the path integral over all the
fields. When $\ep$ is set to zero we denote the expectation value by
$<{\cal O}>$.

We exhibit the position independence for the expectation value of
$\Tr\f^{2}(x) $ which makes the general case obvious. On
differentiating $< \Tr\f^{2}(x)>_{\ep}$ with respect to the point $x$, we find
\bea
d< \frac{1}{8\pi^{2}}\Tr \f^{2}(x) >_{\ep}
&=& <\frac{1}{4\pi^{2}} \Tr \f (x) d_{A}\f(x) >_{\ep} \nonumber \\
&=& \int_{\Phi} \, i \Tr \f (x) \frac{\d}{\d A} \exp{ \left(
\frac{1}{4\pi ^{2}} \int_{\S} \Tr i\f F_{A}
+ \frac{\ep}{8 \pi^{2}} \int_{\S} d\m \Tr \f^{2} \right)} \nonumber
\\
&=& 0 \, . \label{xind}
\eea
In
the last line we have used the fact that the path integral over a total
divergence in function space is zero. To establish that (\ref{pob}) is
independent of the points $x_{i}$ simply requires repeated use of the
reasoning employed in (\ref{xind}). This type of invariance, observables
not depending on the local structure of the underlying manifold, is a
hall-mark of a topological field theory.

The above derivation, however, needs some qualification. Firstly we have not
specified any gauge fixing. Choosing the covariant gauges, for example,
will alter the $A$ variational equation. However, as we are calculating
expectation values of gauge invariant quantities, we do not expect gauge
fixing to alter the picture. Indeed  it is not too difficult to show
that in the Landau gauge, even though the $A$ equation of motion is
altered, (\ref{xind}) formally holds. Standard arguments then extend
this result to all gauges.

The second point to take note of is the fact that we have neither regularised
nor renormalised to which we now turn.

\noindent \underline{Standard Renormalisations}

Yang-Mills theory in two-dimensions is super-renormalisable and there
are no ultraviolet infinities associated with diagrams involving
external gauge $A$ or $\f$ fields. Furthermore, as we are concentrating
on compact manifolds, there are no infrared divergences associated with
these diagrams either. However, because of the coupling of $\f$ to the metric,
there are diagrams which do not involve external $\f$ or $A$ legs but do
have external background graviton legs and which require regularisation.
These terms arise in the determinants that are being calculated and they
depend only on the area and topology of $\Sg$, that is they have the
form
\be
 \a_{1}(\ep) \int_{\Sg} d\m \, + \, \a_{2}(\ep)\int_{\Sg}\frac{R}{4\pi} \,
, \label{st}
\ee
which may be termed area and topological standard renormalisations
(the integral in the second term is just the Euler number $\c(\S)=
2-2g$ 0f $\S$). We wish
to ensure that the scaling invariance that allowed us to move
all of the metric dependence into $\ep$ is respected, accordingly only
those regularisation schemes which preserve this symmetry are to be
considered. Thus the dependence of $\a_{1}$ on $\ep$ is
fixed to be $\a_{1}(\ep) = \ep \beta$ and $\a_{2}$ and $\beta$ are
independent of $\ep$.

\subsection{A Topological Gauge Theory in Two Dimensions}

When we set $\ep=0$ in the action the theory, as we saw, does not have
any explicit dependence on the metric. This means that, providing no
other metric dependence creeps in (and it won't), the theory based on
this action is a topological field theory. The action in this case is
simply $i \Tr \f F_{A}$ and is an example of what are known as $BF$
theories. The partition function of a $BF$ theory, in good situations,
yields the volume of the moduli space of flat connections with the
volume form given by the Ray-Singer Torsion.

Indeed, in line with the general theory, the two dimensional model yields a
volume for the space of flat connections on the Riemman surface
$\S$. At this value of $\ep$ (\ref{pi}) becomes
\be
\int DA D\f \exp{ \left( \frac{1}{4\pi ^{2}} \int_{\S} \Tr i\f F_{A}
 \right)} = \int DA \d \left( F_{A} \right)
= Vol \left(\cal M_{{\cal F}}\right) \, , \label{vol}
\ee
where ${\cal M_{{\cal F}}}$ denotes the space of flat connections.
Clearly, from the second equality in (\ref{vol}), the path integral is
giving us a volume of ${\cal M_{{\cal F}}}(\S,G)$.

The space ${\cal M_{{\cal F}}}$ is known to be an orbifold, that is, a
manifold except at certain singular points. At a singular point it looks
like a `cone' with the singular point at the apex. As far as volumes are
concerned the singular points should pose no problems (the volume of a
cone is not altered if we excise the apex). The only problem we are
faced with then, at the moment, is to decide what volume is being
calculated. Are we calculating the Riemannian volume of ${\cal M_{{\cal
F}}}$ with some preferred metric, with five times that metric, or some
other volume? Unfortunately there is no obvious volume form appearing in
the formula above, as the Ray-Singer Torsion is ``trivial'' in even
dimensions, so it is difficult, at this point, to answer the question.

Fortunately, Witten has answered this question for us, we are calculating
the symplectic volume of ${\cal M_{{\cal F}}}$ with respect to a natural
symplectic two-form on the space of all connections ${\cal A}$. The
point is that a careful analysis shows that the Ray-Singer Torsion does
indeed define a volume for $\MF$ which agrees with the symplectic
volume. There is also a more field theoretic way to see this but for
that we will need a supersymmetric version of (\ref{pi}).

\noindent \underline{Symplectic Structure Of ${\cal A}$}

A symplectic form $\o$ on a $2m$-dimensional manifold is a closed and
non-degenerate two-form,
\be
d\o = 0 \, , \;\;\;\;\;\;\; \det{ \o_{\m \n}(x)} \neq 0 \, ,
\ee
with $\o = \frac{1}{2}\o_{\m \n }(x)dx^{\m}dx^{\n}$. There is a natural
symplectic
form on the space of connections ${\cal A}$ which is inherited from the
Riemman surface $\Sg$. If $\d A_{1}$ and $\d A_{2}$ are tangent vectors
to $A \in {\cal A}$, that is, $\d A_{i} \in \O 1 \Sg$ then the
symplectic form is
\be
\Om (\d A_{1},\d A_{2}) \,= \, \frac{1}{8 \pi^{2}} \int_{\Sg} \, \Tr \d
A_{1} \wedge \d A_{2} \, . \label{symp}
\ee
As $\Om(\, , \, )$ is independent of the point $A \in {\cal A}$ at which
it is evaluated it is closed,
\be
\frac{\d }{\d A} \Om \, = \, 0 \, ,
\ee
and invertibility is clear.

For a finite dimensional symplectic manifold $M$, of dimension $2m$ and
symplectic two-form $\o$, the symplectic volume is given by
\be
\int_{M} \, \frac{\o^{m}}{m!} \, . \label{sympvol}
\ee
This we may express in terms of a Grassman integral as
\be
\int_{M} \, d^{2m} \p \, \exp{ \left( \frac{1}{2} \p^{\m} \o_{\m \n}
\p^{\n} \right)} \, , \label{fdsym}
\ee
which we could also write as $\int_{M} \exp{ \o}$. One notices here that
the $\p^{\m}$ play the role of the basis one-forms $dx^{\m}$, that is,
$\o = \frac{1}{2}\o _{\m \n}dx^{\m} dx^{\n} \sim \frac{1}{2} \o_{\m
\n}\p^{\m} \p^{\n}$.

\noindent \underline{Supersymmetric Extension}

With the finite dimensional example (\ref{fdsym}) in mind we introduce
into the action of (\ref{pi}) the symplectic two form (\ref{symp}). The
new action is
\be
S = \frac{i}{4\pi^{2}}\int_{\Sg} \Tr\left(  \f F_{A}  + \frac{1}{2} \p
\p  \right) + \frac{\ep}{8\pi^{2}}\int_{\Sg} d\m \Tr \f^{2} \, , \label{sbf1}
\ee
where, by an abuse of notation, the fields $\p \in \O 1 \Sg$ are understood
to be Grassman
representatives of one-forms on ${\cal A}$. Correspondingly the path integral
becomes
\be
Z_{\Sg}(\ep) =\int DA D\f  D\p \exp{\left(\frac{i}{4\pi^{2}}\int_{\Sg}
\Tr\left(
\f  F_{A}  + \frac{1}{2} \p
\p  \right) + \frac{\ep}{8\pi^{2}}\int_{\Sg} d\m \Tr \f^{2} \right)} \, .
\label{ymp}
\ee
At this point we have simply producted our original path
integral with the path integral over the field $\p$ but this makes for a
great interpretational improvement as we will see shortly. We have kept
$\ep \neq 0$ in (\ref{ymp}), which seems to go against the complete
metric independence which one would expect in a topological theory, for
later use.

The action (\ref{sbf1}) is invariant under the following supersymmetry
transformations
\be
\d A = \p \, , \; \; \; \; \d \p = d_{A} \f \, , \; \; \; \;  \d \f =0
\, . \label{sup}
\ee
With our understanding that the $\p$ represent elements of $\Om^{1}({\cal
A})$ we see that $\d$ acts like exterior differentiation. However,
$\d$ does not square to zero and one finds instead that
\be
\d^{2} = {\cal L}_{\f} \, ,
\ee
with ${\cal L}_{\f}$ being a gauge transformation with gauge
parameter $\f$. On functions and forms that are gauge invariant $\d$
does square to zero and by restricting ones attention to such objects
one is said to be working `equivariantly'.

Notice that both $\p$ and $\f$ are playing dual roles here. In the above
notation, $\p \in \O 1 \Sg$ but, as $\p$ is Grassman valued, we
have $\p \in \Omega^{1}({\cal A})$. It is more correct then to think of $\p$ as
a two-form, as simultaneously a one-form on $\Sg$ and a one-form on
${\cal A}$. One says that $\p$ is a $(1,1)$-form. A similar duality
holds for $\f$ as we have both $\f \in \O 0 \Sg$ and by the second
equation of (\ref{sup}) $\f \in \Omega^{2}({\cal A})$. In this case $\f$ is
said to be a $(0,2)$-form. To complete the dictionary we note that
$F_{A}$ is a $(2,0)$-form. Lumping the various two forms together
suggests that they are components of a `universal' two-form which is
indeed the case. To see why these geometric structures come out the way
they do one should consult e.g~the review \cite{pr}.

One consequence of the supersymmetry (\ref{sup}) is that we may easily
re-establish that
expectation values of products of invariant polynomials of $\f$ do not
depend on the points at which they sit. For example,
\bea
d <\frac{1}{8 \pi^{2}} \Tr \f^{2}(x)>_{\ep} &=& <\frac{1}{4 \pi^{2}} \Tr [d\f
(x)] \f(x)>_{\ep} \nonumber \\
&=& <\frac{1}{4 \pi^{2}} \Tr [d_{A} \f(x)] \f(x)>_{\ep} \nonumber \\
&=& <\frac{1}{4 \pi^{2}} \Tr \d [\p(x)  \f(x)] \, >_{\ep} \nonumber \\
&=& 0 \, .
\eea
The last line follows by supersymmetric invariance. From this we
conclude that
\be
d <\prod_{i=1}^{n}\frac{1}{8 \pi^{2}} \Tr \f^{2}(x_{i})>_{\ep} = 0 \, ;
\ee
here $d$ stands for differentiation at any of the points $x_{i}$. This
is in agreement with (\ref{xind})\footnote{The Schwinger-Dyson and
supersymmetry proofs of position independence are almost identical. The
supersymmetry path is sometimes easier to follow.}. Similar reservations
to those voiced in the paragraph after (\ref{xind}) need to be
re-iterated here with an additional point related to the supersymmetry,
namely that gauge fixing may well spoil the supersymmetry of the theory.
There are two ways out. The most extensively used is to combine the
supersymmetry
and the BRST symmetry of the gauge fixing procedure into one overall
symmetry. The second would be to work in a gauge which preserves the
supersymmetry from the outset. We will use a hybrid of the two in the
following so that arguments of the type employed above remain valid.

\noindent \underline{Topological Observables}

As the expectation values of these products do not depend on the points of
$\Sg$ we may average over the entire manifold without changing the
results. In equations this means
\be
<\prod_{i=1}^{n}\frac{1}{8 \pi^{2}} \Tr \f^{2}(x_{i})>_{\ep} = <
\left(\frac{1}{8 \pi^{2}} \int_{\Sg} d\m \, \Tr \f^{2} \right)^{n} >_{\ep} \, .
\ee
Furthermore, a glance at the partition function of Yang-Mills theory
(\ref{pi}) or (\ref{ymp}) shows us that
\bea
\frac{\partial^{n} Z_{\Sg}(\ep) }{\partial \ep^{n}} &=& <
\left(\frac{1}{8 \pi^{2}} \int_{\Sg} d\m \, \Tr \f^{2} \right)^{n} >_{\ep}
\nonumber \\
&=& <\prod_{i=1}^{n}\frac{1}{8 \pi^{2}} \Tr \f^{2}(x_{i})>_{\ep} \, .
\label{difz}
\eea

We introduce some notation,
\be
{\cal O}_{0} = \frac{1}{8\pi^{2}} \Tr \f^{2} \, ,
\ee
so that one writes (\ref{difz}) as
\be
\frac{\partial^{n} Z_{\Sg}(\ep) }{\partial \ep^{n}} = < \prod_{i=1}^{n}
{\cal O}_{0}(x_{i}) >_{\ep} \, .
\ee

For the topological theory $\ep$ may be thought of as a an arbitrary
parameter that has nothing to do with the area of the manifold $\Sg$.
With this interpretational change the Yang-Mills path integral becomes a
generating functional for the topological theory. In the standard way,
one differentiates $Z_{\Sg}( \ep)$ a number of times with respect to
$\ep$ and then evaluates at $\ep =0$. In order to deal with the other
polynomial invariants, in this way, one introduces into the action
$\ep_{m}\int_{\Sg} d\m \, P_{m}$ though we shall not be concerned with
this generality here.

The ${\cal O}_{0}$ form one of the basic sets of topological
observables. There are two more sets of topological observables that one
may construct. Before we exhibit these, let us note what our criteria
for an observable is. We certainly want our observables to be metric
independent and invariant under the supersymmetry. However, if the
observables are supersymmetric variations of something ($\d$ exact), then
their expectation values will vanish. So we are searching for observables
which are metric independent, supersymmetric but not $\d$ exact.

Another interesting set of observables that fulfills the criterion are
\be
{\cal O}_{1}(\gamma) = \frac{1}{4\pi^{2}}\int_{\gamma} \Tr ( \p \f) \, ,
\ee
where $\ga$ is any one-cycle of $\Sg$. Supersymmetry invariance is
almost immediate,
\be
 \d \int_{\gamma} \Tr ( \p \f) =
- \int_{\gamma}d \Tr \f^{2} = 0 \, .
\ee
Just as we showed that expectation values of $\Tr \f^{2}$ do not depend
on the points at which they are evaluated one may establish that the
expectation value of $\int_{\gamma} \Tr ( \p \f)$ depends
only on the homology class of $\gamma$. Add to $\gamma$ a homologically
trivial piece $\d \gamma = \partial \Gamma$, then
\bea
 \int_{\gamma + \d \gamma} \Tr ( \p \f) - \int_{\gamma} \Tr ( \p \f)
&=&
\int_{\d \gamma} \Tr ( \p \f) \nonumber \\
& =& \int_{\Gamma } d\Tr ( \p \f) \nonumber \\
&=& -\d  \int_{\Gamma} \Tr (\f F + \p \p) \, ,
\eea
and the expectation value of the last term will vanish by supersymmetry
invariance.

Expectation values of these observables can also be obtained by
differentiating the Yang-Mills partition function. We wish to calculate
\bea
< \prod_{i=1}^{n}{\cal O}_{1}(\gamma_{i})>_{\ep}
 & = & \int DA \, D\p \, D\f \exp{\left(
\frac{i}{4\pi^{2}} \right.  \int_{\Sg} \Tr\left(  \f F  + \frac{1}{2} \p
\p  \right) } \nonumber \\
& & \; \; \; \; \; \; \; \; \; + \left. \frac{\ep}{8\pi^{2}}\int_{\Sg} d \m\Tr
 \f^{2}
\right) . \prod_{i=1}^{n} \frac{1}{4\pi^{2}} \oint_{\gamma_{i}} \Tr \f \p
  \, . \label{inta}
\eea
Here $n$ must be even or this vanishes because the action is invariant under
$\p
\rightarrow - \p$ while the integrand changes sign if $n$ is odd. A simple
way to perform this integral is to introduce $n$ anti-commuting
variables $\e_{i}$ and consider instead the partition function
\bea
Z_{\Sg}(\ep, \e_{i})& =& \int DA \, D\p \, D\f \, \exp{\left(
\frac{i}{4\pi^{2}} \right.  \int_{\Sg} \Tr\left(  \f F  + \frac{1}{2} \p
\p  \right)} \nonumber \\
& & + \left. \frac{\ep}{8\pi^{2}}\int_{\Sg} d\m \Tr \f^{2} +
\frac{1}{4\pi^{2}} \sum_{i=1}^{n}  \e_{i} \oint_{\gamma_{i}} \Tr \f \p
\right) \, . \label{int1a}
\eea
On differentiating this with respect to each of the $\e_{i}$ (in the
order $i=n$ to $i=1$) and then setting these Grassman variables to zero
one obtains (\ref{inta}). Now we introduce De Rham currents $J$ with the
following properties
\be
\int_{\Sg} J(\gamma_{i}) \Lambda = \oint_{\gamma_{i}} \Lambda \, , \;
\; \; dJ = 0 \,
\ee
for any one form $\Lambda$. One completes the square in (\ref{int1a}) in
the $\p$ field
\be
\p \rightarrow \p - i\sum_{i=1}^{n} \e_{i}J(\gamma_{i}) \f \, ,
\ee
to obtain
\bea
Z_{\Sg}(\ep, \e_{i})& =& \int DA \, D\p \, D\f \exp{\left(
\frac{i}{4\pi^{2}} \right.  \int_{\Sg} \Tr\left(  \f F  + \frac{1}{2} \p
\p  \right) } \nonumber \\
& & + \left. \frac{\ep}{8\pi^{2}}\int_{\Sg}d\m \Tr \f^{2}  -
\frac{i}{4\pi^{2}} \sum_{i<j}^{n}  \e_{i} \e_{j}\int_{\Sg}
J(\gamma_{i})J(\gamma_{j}) \Tr \f^{2} \right) \, . \label{int2a}
\eea
The terms with $i=j$ vanish as $\e_{i}^{2}=0$, so that there are no
problems with self intersections.
The De Rham currents have delta function support onto their associated
cycles so that, for any zero form $\Psi$, ($i \neq j$)
\be
\int_{\Sg} J(\gamma_{i})J(\gamma_{j}) \Psi = \sum_{P \in \gamma_{i}
\cap \gamma_{j}} \sigma(P) \Psi(P) \, ,
\ee
with $P$ the points of intersection of $\gamma_{i}$ and $\gamma_{j}$
and $\sigma(P)$ ($= \pm 1$) the oriented intersection number of
$\gamma_{i}$ and $\gamma_{j}$ at $P$. This means that, in the path integral,
\bea
& & \frac{1}{4\pi^{2}} \sum_{i<j}^{n}  \e_{i} \e_{j}\int_{\Sg}
J(\gamma_{i})J(\gamma_{j}) \Tr \f^{2} \nonumber \\
& =& \frac{1}{4\pi^{2}} \sum_{i<j}^{n}
\e_{i} \e_{j} \gamma_{ij}  \Tr\f^{2}(P)
\nonumber \\
& =& \frac{1}{4\pi^{2}} \sum_{i<j}^{n}  \e_{i} \e_{j} \gamma_{ij}
\int_{\Sg} \Tr \f^{2} \, ,
\eea
where we have used the fact that $\Tr\f^{2}$ does not depend on the point
at which it is evaluated and $\gamma_{ij}= \# (\gamma_{i} \cap \gamma_{j})$
is the matrix of oriented intersection numbers. Putting all the pieces together
we arrive at
\be
Z_{\Sg}(\ep,\e_{i}) = Z_{\Sg}(\hat{\ep}) \, , \label{triv}
\ee
with
\be
\hat{\ep} = \ep - 2 \sum_{i<j} \e_{i}\e_{j}\gamma_{ij} \, .
\ee

For $n=2$ we obtain
\bea
< {\cal O}_{1}(\gamma_{1}) {\cal
O}_{1}(\gamma_{2}) >_{\ep} & =& \frac{\partial}{\partial
\e_{1}}\frac{\partial}{ \partial \e_{2}} Z_{\Sg}(\ep
-2\e_{1} \e_{2} \gamma_{12}) \nonumber \\
&=& 2 \gamma_{12} \frac{\partial}{\partial \ep } Z_{\Sg}(\ep) \, . \label{ob2}
\eea
Likewise for higher values of $n$ the expectation values of
$ {\cal O}_{1}(\gamma_{i}) $ are obtained on differentiating $Z_{\Sg}(\ep)$.

Clearly, expectation values of mixed products
\be
< \prod_{i=1}^{k}{\cal O}_{0}(x_{i}) \prod_{j=1}^{n} {\cal
O}_{1}(\gamma_{j})  >_{\ep} \, , \label{genobs}
\ee
are similarly obtained.

The last of the observables of interest to us is the topological action
itself
\be
{\cal O}_{2} = \frac{i}{4\pi^{2}}\int_{\Sg} \Tr \left( i \f F_{A} +
\frac{1}{2}  \p \wedge \p \right) \, .
\ee

\noindent \underline{Interpretation}

As the path integral $Z_{\Sg}(0)$ essentially devolves to an integral
over $\MF$, $\p$ and $\f$ should be thought of as elements of
$\Omega^{1}(\MF)$ and $\Omega^{2}(\MF)$ respectively. The observables we
have been considering are $\d$ closed (invariant under the
supersymmetry) but not $\d$ exact, so they should descend to
elements of the cohomology groups of $\MF$. More should happen - they
should generate the cohomology, but, at the level that we are
working at, we can only give a heuristic argument for this: from the path
integral point of view there are simply no other topological observables that
we can write down so the ones we have should (we hope) encode all the
relevant information.

The (rational) cohomology classes are the observables ${\cal O}_{i}$
\bea
 \frac{1}{8 \pi^{2}}\Tr \f^{2} & \in & H^{4}\left(\MF (\Sg , \bG) \right) \, ,
\nonumber \\
 \int_{\ga} \Tr
\p \f & \in & H^{3}\left(\MF (\Sg , \bG ) \right) \, , \nonumber \\
 \frac{i}{4\pi^{2}}\int_{\Sg} \Tr \left( i \f F_{A} + \frac{1}{2} \p
\wedge \p \right) & \in & H^{2}\left(\MF (\Sg , \bG ) \right) \, .
\eea
There is no invariant $(0,1)$ form that we can write down so we conclude
that $H^{1}\left(\MF (\Sg , SU(n)) \right)$ is trivial. For $U(n)$, on the
other
hand, one can construct elements of $H^{1}\left(\MF (\Sg) \right)$
namely $\int_{\ga}\Tr \p$. The
higher invariant polynomials $P_{n}$ would represent elements of
$H^{2n} \left(\MF (\Sg , \bG) \right)$.

On any $n$ dimensional manifold we may integrate an
$n$-form without the need to introduce a metric. The moduli space has
dimension (for $g>1$) $(2g-2) \dim \bG$ so that any product of the
observables as in (\ref{genobs}) with $4k + 3n = (2g-2) \dim \bG$ is a
form that may be integrated on $\M$. On the other hand, once the
constraint that $F_{A}=0$ has been imposed, the path integral over
${\cal A}$
devolves to an integral over $\MF$. In this way (\ref{genobs}) is seen to
be the integral over $\MF$ of a $(2g-2) \dim \bG$-form. Let us denote with
a hat the differential form that an observable corresponds to. Then
(\ref{genobs}) takes the more suggestive form
\be
< \prod_{i=1}^{k}\frac{1}{4\pi^{2}}{\cal O}_{0}(x_{i}) \prod_{j=1}^{n
}\frac{1}{4\pi^{2}}{\cal O}_{j}(\gamma_{1}) > = \int_{{\cal M}_{F}}
\prod_{i=1}^{k}\frac{1}{4\pi^{2}}\hat{{\cal O}}_{0}(x_{i}) \prod_{j=1}^{n
}\frac{1}{4\pi^{2}}\hat{{\cal O}}_{j}(\gamma_{1}) \exp{\Omega} \, .
\label{rel}
\ee
When $4k +3n = (2g-2) \dim \bG$, the symplectic form makes no
contribution. However, if $4k +3n = 2m < (2g-2) \dim \bG$ there will also
be contributions from the action to soak up the excess form-degree.
On expanding the exponential, the symplectic form $\Omega(\p,\p)$ raised to
the power $(g-1)\dim \bG -m$ will survive the Grassman integration.

{}From (\ref{triv}) we are able to conclude that, up to numerical
constants, even powers of the classes generating $H^{3}(\MF)$ can be replaced
by the appropriate powers of generators of $H^{4}(\MF)$. This means that
in (\ref{rel}) we need only consider products of ${\cal O}_{0}$.

In this discussion we have presumed that $\MF$ is a reasonably nice
space. For the group $SO(3)$ this is the case and in performing the above
path integrals the correspondence (\ref{rel}) holds. Integrals over
products of non-trivial cohomology classes give direct information about
intersection numbers on the moduli space.
It turns out that $\MF(\Sg, SU(2))$ is not such a
nice space. An indication of this, that we will see, is that the
Yang-Mills partition function has a non-analytic behaviour in $\ep$ as
$\ep$ approaches zero.

\noindent \underline{Standard Renormalisations?}

The picture that we have obtained needs to be tempered by the possibility
of a standard renormalisation of the form
\be
\ep \beta \int_{\Sg} \, d\m \, . \label{stren}
\ee
The appearance of such a term would mean that we would need to redefine our
cohomology classes $\Tr \f^{2}$. However, the method we employ to solve
the theory is in agreement with the derivation based on fixed point
theorems. This means that we will not need to make such a standard
renormalisation. Consequently the `naive' considerations above do not need to
be altered.

\subsection{Weyl Integral Formula for Lie Algebras}

In order to proceed we must fix on a gauge.
There is a useful choice of gauge that makes the task of evaluating
(\ref{pi}) particularly simple. The preferred gauge choice is well known
to physicists, it is the unitary gauge used to elucidate the particle
content of spontaneously broken gauge theories. This is a condition
imposed not on the gauge connection but, rather, on the Higgs field, which
in our context is $\f$ appearing in (\ref{pi}).

This gauge amounts to setting $\f^{\lk} = 0$, a condition that can certainly
always be imposed pointwise, i.e.~at the level of finite dimensional Lie
algebras. It can also be imposed locally, but there may be obstructions
to implementing it globally via {\em continuous} gauge transformations. We will
come back to this and its consequences, which need to be carefully kept
track of, below - see the comments in \cite{btver} as well as \cite{btwif}
for a detailed treatement of this issue.

In any case, continuing for the time being to treat this as an ordinary
gauge condition, we note that it enforces a partial gauge
fixing preserving the Cartan subalgebra and the corresponding Abelian
gauge symmetry. Later we will fix this residual symmetry by imposing the
Landau gauge condition on the gauge fields lying in this subalgebra. Our
presentation below, which essentially imposes the two conditions
independently, misses cross terms amongst the ghosts. These cross terms
may, in any case, be shown not to contribute and we lose nothing by not
imposing all the gauge conditions at once. This gauge does not involve
the connection so that our arguments which made use of the
Schwinger-Dyson equation remain valid. Likewise, as $\f$ does not
transform under the supersymmetry transformations, this gauge preserves
that symmetry and any inferences based on it.

We turn to a quick review of the Lie algebra theory that we will need in
order to see what the unitary gauge means for us.

\noindent \underline{The Unitary or Torus Gauge}

We note that that it is a theorem that all elements $\f \in \lg$ can be
conjugated into the Cartan subalgebra. This means that there exist $g(\f)
\in \bG$ such that $g(\f)^{-1} \f g(\f) \in \lt$. As we can conjugate
any element $\f \in \lg$ into $\lt$,
we have found that for conjugation invariant theories we are allowed to
gauge fix $\f$ to lie in $\lt$ (modulo the above {\em caveat} about
possible topological obstructions to achieving this gauge globally).
Alternatively put,
\be
\f^{\lk}=0 \, ,
\ee
is an allowed gauge (here the superscript denotes the components of $\f$
in $\lk$). Yang-Mills theory is an example of a theory that is,
pointwise, conjugation invariant. This means that we may set $\f^{\lk}$
to zero pointwise.

\noindent \underline{Weyl Integral Formula for Lie Algebras}

There is a very beautiful formula in Lie group theory, due to Weyl, that
allows one to express the integral over the group, of a class function, as
an integral over the maximal torus. This will be exhibited in section 3.5, here
we will derive an analogous formula for integrals over the Lie algebra
$\lg$ of $\bG$.

Let $f(\f)$ be an $\Ad$ invariant function on $\lg$, i.e.
\be
f(g^{-1}\f g) = f(\f) , \; \; g \in \bG ,
\ee
and also take $f$ to be integrable. Think of $\lg$ as $\RR^{n}$
for an appropriate $n$ and let the measure on $\lg$ be the
standard Riemann-Lebesgue measure, $d^{n}\f/\sqrt{2\pi}^{n}$. Our aim is
to derive a formula
for the integral of $f$ over $\lg$, with this measure, in terms of an
integral over $\lt$ with a somewhat different measure. This is done
using the Faddeev-Popov procedure.

The ``modern'' rule here is to define a
nilpotent BRST operator $Q$ and then add a $Q$ exact term to the
``action'' which at once fixes the gauge and adds the appropriate
Faddeev-Popov ghost term. In the case at hand, with
\be
\f = i \sum_{i=1}^{r} \f_{i} t_{i} + i \sum_{\a} \f_{\a} E_{\a} \, ,
\ee
we define
\bea
Q \f_{\a} = \a(\f) c_{\a}  \, , \;\; \;   Qc_{\a}& =& 0 \, ,
\nonumber \\
 Q \f_{i} = 0 \, , \; \; \; Q \bar{c}_{\a} = b_{\a} \, ,  \;\; \; & &  Q
b_{\a} = 0 \, .
\eea
Here $\a(\f) = \a(t_{i}) \f_{i}$ and the $\f_{i}$ are real while
$\f_{\a}^{*} = \f_{-\a}$. With these rules it is apparent that $Q^{2}=0$.
We introduce into the integral an exponential term with exponent
\be
\{ Q, i\sum_{\a} \bar{c}_{-\a} \f_{\a} \} = i\sum_{\a} b_{-\a} \f_{\a} - i
\sum_{\a} \a(t_{i}) \bar{c}_{-\a} \f_{i} c_{\a} \, ,
\ee
which clearly gives a delta function constraint onto the Cartan
subalgebra. So up to a universal constant
\bea
\int_{\lg} f(\f) &=& N \int_{\lg} \exp{\left(i \sum_{\a} b_{-\a} \f_{\a} -
i\sum_{\a} \a(t_{i}) \bar{c}_{-\a} \f_{i} c_{\a}  \right)} f(\f) \nonumber \\
&=& N \int_{\lt} f(\f^{\lt}) \det{}_{\lk} (\ad(\f^{\lt})) \, . \label{int}
\eea

The normalization constant $N$ is not unity. The reason for this is that
there is still a subgroup that maps the Cartan subalgebra to itself
nontrivially by conjugation. This, finite group $W$, is called the Weyl
group. The normalisation constant is then the inverse of the order of $W$;
$N  = 1/\mid W \mid$. To see that this is correct suppose that $g \in
\bG$ conjugates $\f \in \lg$ to the element $\f' \in \lt$, then $gw$
will map $\f$ to $w^{-1}\f'w \in \lt$ for all $w \in W$. We cover the
Cartan subalgebra $\mid W \mid$ times in this way.

\noindent \underline{Gauge Fixing Again}

The reader is perhaps perplexed by the above procedure. We have used
infinitesimal transformations to land on the gauge $\f^{\lk}=0$ and to
generate the ghost terms, while one knows that to achieve the gauge
choice, group conjugation needs to be employed. To see that,
nevertheless, the above analysis is correct we rederive the integral
formula (\ref{int}) in the ``old fashioned'' way. Let
\be
\int_{G} \d\left([g^{-1}\f g]^{\lk}\right) \Delta(\f) \, = 1 \, . \label{fp}
\ee
The delta function is a Dirac delta function (on $\lk$) and the group measure
is
Haar with $\int_{G} =1$. The Faddeev-Popov determinant $\Delta$ is defined
by this equation. One property of $\Delta$ that we will use momentarily
is that $\Delta(h^{-1}\f h) =\Delta(\f)$, an easy consequence of the
invariance of the group measure.

Now we have
\bea
\int_{\lg} f(\f) \, &=& \int_{\lg} \int_{G} \, f(\f)
\d\left([g^{-1}\f g]^{\lk} \right) \Delta(\f) \nonumber \\
&=& \int_{\lg} \, f(\f)
\d\left(\f^{\lk} \right) \Delta(\f) \nonumber \\
&=& \int_{\lt} f(\f^{\lt}) \Delta(\f^{\lt}) \, .
\eea
The first equality follows by inserting unity (\ref{fp}), the second by
conjugating $\f \ra g\f g^{-1}$ and using the invariance properties of
$f$ and $\Delta$, and the last by integrating over the $\f^{\lk}$
components (by the delta function). We see that we only need to
determine $\Delta(\f^{\lt})$ to get the required formula.

{}From (\ref{fp}) we have
\be
\Delta(\f^{\lt})^{-1} = \int_{G}
\d\left([g^{-1}\f^{\lt}\lt g]^{\lk}\right) \, .\label{rhs}
\ee
Any group elements that lie in the maximal torus will factor through.
This means that the delta function has its support around the (chosen)
maximal torus, so we can expand $g \approx g_{t}(1 + \l)$ for $\l \in
\lk$ (and up to the action of the Weyl group - this will give rise to
a  factor $|W|$ on the right hand side of (\ref{rhs})).
One now sees why the previous derivation only required Lie algebra
data; one does use the group to conjugate into the Cartan subalgebra,
but to determine the correct measure around this slice local
deformations suffice. Remembering the correct normalization factor,
we find in this way that
\be
\Delta(\f^{\lt}) = \trac{1}{|W|} \det{}_{\lk} [\ad(\f^{\lt})] \, .
\ee

\noindent \underline{Cautionary Remarks}

This mapping of an integral over the whole Lie algebra to an integral
over a Cartan subalgebra, while useful, must be handled with some
caution. We show the point with the following integral
\be
\int \frac{d^{n} \f}{\sqrt{2\pi}^{n}} \int \frac{d^{n}
\ga}{\sqrt{2\pi}^{n}}  \exp{\left( i \Tr \f \ga \, +\frac{\ep}{2}  \Tr \f^{2}
+ \frac{\ep'}{2} \Tr \ga^{2}
\right)} \, = \, \left(\frac{1}{1 -\ep \ep'}\right)^{n/2} \, . \label{int1}
\ee
One may set either $\ep$ or $\ep'$ (or both) to zero in this
equation and it clearly makes sense.

Let us use the Weyl integral
formula with $\ep'= 0$ and with the rotation of $\f$ to $\f^{\lk}=0$. We
run into  a problem directly. The formula obtained in this way is zero times
infinity. The infinity comes from the fact that $\ga^{\lk}$ no longer
makes an appearance in the exponent so that one finds for the
$\ga^{\lk}$ integral $\infty^{n-r}$. The zero comes from the fact that
the $\ga^{\lt}$ integral imposes $\f^{\lt}=0$ so that the determinant
$\det{}_{\lk}{\left( \ad \f^{\lt} \right) }$ yields $0^{n-r}$. Re-instating
the term $\ep' \Tr \ga^{2}$ into the exponent is a way of regulating the
problem and taking the limit $\ep' \rightarrow 0$ at the end is in
agreement with (\ref{int1}) in that limit. Notice that in this
discussion the term $\ep \Tr \f^{2}$ played no role. One could set $\ep
=0$ with impunity or, put another way, instead of setting $\f^{\lk}=0$
we could have fixed $\ga^{\lk}=0$, without running into any problems.

\subsection{Evaluation of the Partition Function}

We return to our task of evaluating the partition function (\ref{pi}) in
the gauge $\f^{\lk}=0$. In this gauge the action, including the ghost
terms, becomes
\be
\frac{1}{4\pi ^{2}} \int_{\S} \Tr \, i\f^{\lt} F_{A}^{\lt}
+\, \frac{\ep}{8 \pi^{2}}  \int_{\S} d\m \,\Tr \, \f^{\lt} \f^{\lt}  +
\frac{1}{4\pi ^{2}}\Tr \int_{\S} d\m \, (b^{\lk} \f^{\lk} \, + \,
\bar{c}^{\lk} [ \f^{\lt}, c^{\lk} ] ) \, .
\ee

\noindent \underline{Area dependence}

Notice that while we have, unavoidably, introduced area dependence (through
$d\m$) in both the gauge fixing term and the ghost term this dependence
`cancels' out between them. One way of exhibiting this is to note that
the combination of those two terms is $Q$ exact and hence their
variation with respect to the metric is $Q$ exact. Alternatively, scale
the metric by $\lambda$, this sends the area element to
$\lambda^{2}d\m$, follow this by a scaling $b \ra b/\lambda^{2}$,
$\bar{c} \ra \bar{c}/ \lambda^{2}$ so that the combination of scalings
leaves the action (apart from the $\f^{2}$ term) invariant. The
Jacobians of the transformations of the $b$ and $\bar{c}$ fields are
inverses of each other, so overall the Jacobian is unity. With this
understood, we from now on drop all reference to $b^{\lk}$ and
$\f^{\lk}$.

\noindent \underline{The Action}

Let us expand the gauge fields as
\be
A^{\lk} = i\sum_{\a} A^{\a} E_{\a} \, , \;\;\; \; A^{\lt} = i\a_{l} A^{l}
\, ,
\ee
where we impose the reality condition that $A_{\a}^{*}=
A_{-\a}$.\footnote{ On choosing a complex structure on $\Sg$, as we will
do in appendix B this translates into $(A^{\a}_{z})^{*} =
A^{-\a}_{\bar{z}}$.} Likewise we expand the ghost fields in the
same basis
\be
c^{\lk} = i\sum_{\a} c^{\a} E_{\a}\, , \; \; \; \; \bar{c}^{\lk} = i\sum_{\a}
 \bar{c}^{\a} E_{\a} \, ,
\ee
and choose as a reality condition $(c^{\a})^{*} = \bar{c}^{-\a}$. On the
other hand we expand $\f^{\lt}$ in the basis of fundamental weights,
\be
\f^{\lt} = i\f_{l}\lambda^{l}\, .
\ee

In this basis the action is
\bea
\frac{1}{4\pi ^{2}} \int_{\S}& &  \, \left(-i \sum_{l =1}^{r}\f_{l}
dA^{l}  \, + \, \sum_{\a} \, \a(\f)
A^{\a}A^{-\a} \right) \nonumber \\
& - & \frac{\ep}{16 \pi^{2}}  \int_{\S} d\m \,\sum_{l=1}^{r} \, \f_{l}
\f_{l}  + \frac{1}{4\pi ^{2}} \int_{\S} d\m  \,  \sum_{\a} \,
\a(\f)\bar{c}^{-\a} c^{\a}
\eea

We shall first integrate out the gauge field components $A^{\lk}$ and
the ghost fields $\bar{c}^{\lk}$ and $c^{\lk}$. The integration over the
gauge field yields
\be
\Det_{\lk} \left( \ad(\f^{\lt}) \right) _{\Omega^{1}(\S)}^{-1/2} \, ,
\label{det1}
\ee
while that over the ghosts gives
\be
\Det_{\lk} \left( \ad(\f^{\lt}) \right) _{\Omega^{0}(\S)} \, . \label{det0}
\ee
The product of these two determinants is almost unity. We expect this as
a vector is `like' two scalars in two dimensions. More precisely we
have, by the Hodge decomposition theorem, that any one-form $\o$ on a
compact Riemann surface may be uniquely written as the sum of an exact,
a coexact and a harmonic form
\be
\o = d\a \, + \, d^{*}\beta \, + \, \gamma \, ,  \label{hodge}
\ee
where $\a$ and $\beta$ are zero-forms. Here $d^{*}$ is the adjoint of $d$
with respect to the scalar product
\be
(\o_{1}, \o_{2}) = \int_{\S} \o_{1}*\o_{2} \, , \label{in}
\ee
and from $*^{2}= (-1)^{p}$ acting on a $p$-form we deduce that $d^{*} = -
*d*$.

The Hodge decomposition (\ref{hodge}) tells us to what extent the
identification of a one-form with two zero-forms can be made. From
(\ref{hodge}) we see that the zero-form $\a$ enters as $d\a$ so that its
harmonic piece (the constant zero-form) does not enter and likewise for
$\beta$. Denoting the space of harmonic forms by $H^{*}(\S,R)$, we may
figuratively decompose the space of one-forms as
\be
\Om ^{1}(\S,R) =  [\Om^{0}(\S,R) \ominus H^{0}(\S,R)] \oplus
[\Om ^{0}(\S,R) \ominus H^{0}(\S,R)] \oplus H^{1}(\S,R) \,.
\ee

Using the fact that the harmonic modes are orthogonal, with respect to
the inner product (\ref{in}), to all the other modes, we may deduce that
the product of determinants (\ref{det1}) and (\ref{det0}) is more or
less equal to
\be
\frac{\det{}_{\lk}\left( \ad(\f^{\lt}) \right)_{H^{0}(\S)}  }{
\det{}_{\lk}^{1/2}\left( \ad(\f^{\lt}) \right)_{H^{1}(\S)} } \, .
\label{detym}
\ee
If $\f$ were constant then one could combine these two determinants into
the single
\be
\det{}_{\lk}\left( \ad(\f^{\lt}) \right)^{\c (\S)/2}
\, ,
\ee
where the Euler character of $\Sg$, $\c (\Sg)$, is given in terms of
the Betti numbers, $b^{i} = \dim H^{i}(\Sg)$, by $ \c (\Sg)=
2b^{0}-b^{1}$. The Betti numbers are $b^{0}=1$ and $b^{1}=
2g$ and so consequently $ \c (\Sg)= 2(1-g)$. When $\f$ is not
constant there is still a formula that one may derive (which we do in
appendix B), namely
\be
\frac{\det{}_{\lk}\left( \ad(\f^{\lt}) \right)_{H^{0}(\S)}  }{
\det{}_{\lk}^{1/2}\left( \ad(\f^{\lt}) \right)_{H^{1}(\S)} } =
\exp{\left[ \frac{1}{8\pi} \int_{\S} R \sum_{\a}\log{\a(\f)}\right]}
\ee

\noindent \underline{Winding Numbers and Non-Trivial Torus Bundles}

We are now left with a path integral over $\f^{\lt}$ and $A^{\lt}$ to
perform and it is here that the consequences of having implicitly
worked with gauge transformations which are not necessarily continuous (in
order to
achieve the gauge $\f^{\lk}=0$) make their appearance. We briefly
try to explain this point here, referring to \cite{btwif} for full details.

First of all, note that
the $A^{\lt}$ are connections on torus bundles over the Riemann
surface $\S$. Such bundles are completely classified by their first
Chern class, or monopole number. That is, for a given torus bundle, we
have a set of integers $\{ n^{l}, l = 1,\ldots ,r\}$ with
\be
\int_{\S} F_{A}^{l} = 2\pi n^{l} \, . \label{chern}
\ee
As the original $\bG$-bundle we started off with was trivial, we would expect
to have to integrate over torus connections on the trivial bundle only,
provided that everything we did was globally well defined.

On the other hand, there certainly are Lie algebra valued maps $\f$ which
cannot be conjugated into the Cartan subalgebra globally. As an example
(pointed out to us by E.~Witten) consider the map from the two-sphere
$S^{2}$ to $SU(2)$ given by
\be
\f(x_{k}) = \sum_{k} x_{k}\sigma_{k}\;\;,
\ee
where $\sum (x_{k})^{2} =1$ and the $\sigma_{k}$ are Pauli matrices. The
image of this map is a two-sphere in $SU(2)$ and hence it is possible to
associate a winding number $w(\f)$ to $\f$.
As $\f$ is essentially the restriction
to $S^{2}$ of the identity map from $\RR^{3}$ to itself, it is clear that
this winding number is $w(\f)=1$, as can also be inferred from the integral
representation of $w(\f)$ as the integral over $S^{2}$ of the pull-back
of the (normalized) volume form $\omega$ on the target-$S^{2}$,
\be
w(\f) = \int_{S^{2}}\f^{*}\omega =
\trac{1}{4\pi} \int_{S^{2}}\f [d\f,d\f] = 1\;\;.
\label{wind}
\ee
But from this expression it is clear that if $\f$ could be conjugated into
the Cartan subalgebra by a globally well-defined map (an operation under which
its winding number should not change), $w(\f)$ would have to be zero
because the integrand would vanish identically in this gauge.

It can, moreover, be checked that the (now necessarily discontinuous)
gauge transformation mapping $\f$ into the Cartan subalgebra transforms
the torus component of the $su(2)$ gauge field into a connection on a
$U(1)$ bundle with Chern class 1. The upshot of this is that, in general,
this choice of gauge will engender a sum over all isomorphism classes of
torus bundles and that hence in the path integral we have to sum
over all of the $r$-tuples of integers appearing in (\ref{chern}).

\noindent \underline{Integration over Non-Trivial Torus Bundles}

There are various ways of taking (\ref{chern}) into account, one of which
is described in \cite{btym} and \cite{btver}. Here, however, we follow a
different route. In order to deal with (\ref{chern}) one
splits the gauge field $A^{\lt}$ into a classical $A_{c}$ and a quantum
part $A^{q}$. The quantum part is a torus valued one-form while the
classical piece may be taken to satisfy
\be
dA_{c}^{l} = 2\pi n^{l} \o \, ,
\ee
and clearly obeys (\ref{chern}).

The path integral over the torus gauge field can be easily performed.
The relevant part of action is
\be
\sum_{l=1}^{r} [\int_{\S}  \, n^{l} \f_{l} \o \, + \, \int_{\S} \f^{l}
dA^{q}_{l}] \, ,
\ee
which still requires a gauge fixing condition on $A^{q}$. Let us choose
the Landau gauge
\be
d*A^{q}=0 \, ,
\ee
and introduce this into the action together with the usual ghost terms.
The required action is
\be
\sum_{l=1}^{r} [\int_{\S}  \, n^{l} \f_{l} \o \, + \, \int_{\S} \f^{l}
dA^{q}_{l} + \int_{\S}  b^{l} d*A^{q}_{l} + \int_{\S}
\bar{c}^{l} d*d c_{l}] \, .
\ee

The integral over $A^{q}$ yields a delta function constraint,
\be
d\f^{l} + *d b^{l} =0 \, , \label{inst}
\ee
which implies that
\be
d\f^{l} = 0 \, , \; \; \; \; *db^{l}=0 \, .
\ee
To see this, let us take the square of (\ref{inst}) for each $l$,
\bea
0 &=& \int_{\S} (d\f + \, *db)*(d\f + *db) \nonumber \\
  &=& \int_{\S} d\f * d\f \, + db * db \nonumber \\
  &=& (d\f,d\f) + (db, db) \, .
\eea
The right hand side of the last line is a sum of squares so each term must
be zero individually.

As this also implies that the $\f$ equation of motion is unchanged, the
Schwinger-Dyson argument (\ref{xind}) proving position independence of the
$\f$-correlators
survives the gauge fixing of the residual Abelian symmetry. To see that
also the argument based on supersymmetry is still applicable requires a
little more care. Now the supersymmetry transformations (\ref{sup})
should be supplemented by the BRST transformation so that one has e.g.
\be
\tilde{\d} A = \p + d_{A}c \;\;,\;\;\;\;\;\;\tilde{\d}\p = d_{A}\f + [c,\p]
\;\;.
\ee
Choosing the gauge fixing term in the action to be $\tilde{\d}$-exact
everything can then be seen to go through as above.

At this point the fields $A^{q}, b, \bar{c}$ and $c$ are no longer of
interest. However, we would like to see that they contribute no dynamical
determinants to the partition function. Firstly, notice that the harmonic
one-form modes do not make an appearance here, nor do the the $b^{l},
\bar{c}^{l}$ or $c^{l}$ zero-form harmonic modes. These can be gauge fixed
to zero, and their effect will be to multiply the
partition function by an unspecified constant. Ignoring these harmonic
modes we may exchange $A^{q}_{l}$ by its Hodge decomposition
\be
A^{q}_{l} = d\a_{l} + * d\beta_{l} ,
\ee
with
\be
DA^{q} = D\a D\beta \, \Det [d*d*]_{\Om ^{0}} \, . \label{meas}
\ee
Integration over the $b$ and $\beta$ generates a determinant
$\Det[d*d*]_{\Om^{0} }^{-1}$ cancelling the Jacobian in (\ref{meas}).
The integral over $\a$ now yields a delta function
\be
\d(d*d* \f) = \Det[d*d*]_{\Om ^{0}}^{-1} \d'(\f) \, , \label{del}
\ee
where the prime is meant to indicate that the delta function puts no
constraint on the constant $\f$ mode. The determinant that appears in
(\ref{del}) cancels against the determinant that one obtains on
integrating out $\bar{c}$ and $c$. All the determinants have cancelled
and  we are still dealing with a theory with no dynamical degrees of
freedom.

\noindent \underline{Finite Dimensional Integrals}

On putting all the pieces together, we see that the original path
integral devolves to a product of $r$ simple finite dimensional
integrals. All the fields except $\f$ have been integrated out with the
net effect that $\f$ must be space-time independent and with an
insertion of a finite dimensional determinant $\det{}_{\lk}\left( \ad(\f^{\lt})
\right)^{\c (\S)/2} $. The complete integral is
\be
\prod_{l=1}^{r}\sum_{n_{l}} \int d\f^{l} \, \det{}_{\lk}\left( \ad(\f^{\lt})
\right)^{\c (\S)/2} \, \exp{\left(-i \frac{\f^{l}n_{l}}{2\pi} -
\frac{\ep \f^{2}}{16\pi^{2}}\right) }\, . \label{intaa}
\ee

There are two ways of evaluating this integral, the first brings us
directly into the form found in \cite{ewym,btym} giving the partition
function as a sum of irreducible representations with weight the
dimension of the representation times the exponential of the area times
the quadratic Casimir of the representation.
The second method of solution is in line with an evaluation of the path
integral given in \cite{ew2d} and has the advantage of allowing us to
ascertain the behaviour of the partition function as $\ep$ tends to
zero. Both derivations `fall' onto a fixed point set thus yielding an
alternative proof of the fixed point theorem presented in \cite{ew2d}.

\noindent \underline{Solution $1$}

Notice that the sum over $n_{l}$ yields a periodic delta function on
$\f$,
\be
\prod_{l=1}^{r}\sum_{n_{l}} \exp{\left(i\frac{\f^{l}n_{l} }{ 2\pi}\right)} =
\prod_{l=1}^{r }\sum_{n^{l}} \d(\f^{l} - 4 \pi^{2} n^{l}) \, .
\label{pdf}
\ee

Substituting this expression into (\ref{intaa}) gives us
\be
\prod_{l=1}^{r}\sum_{n_{l}} \, \det{}_{\lk}    \left( \ad(n_{\lt})
 \right) ^{\c (\S)/2} \, \exp{\left(
-\ep \pi^{2} n_{l}^{2}\right) } \label{int2}
\ee
At this point we see that the terms in the sum where
$\det{}_{\lk}(\ad(n_{\lt}))=0$ diverge for $g\geq 2$. Elements
$\lambda \in \lt$ which satisfy $\det{}_{\lk}(\ad(\lambda)) \neq 0$ are
called regular elements. For the moment we will only sum over the
regular $n_{\lt}$ that appear in (\ref{int2}) and return to the thorny
question of the non-regular elements later.

In (\ref{int2}), the sum over the Chern classes may be thought of as a
sum over the weight lattice. That is, one sets $\lambda = \sum _{l} n_{l}
\lambda^{l}$ with the weight lattice given by
\be
\Lambda = \ZZ [\lambda^{1}, \dots , \lambda^{r}] \, .
\ee
In this way we obtain,
\be
\sum_{\lambda}  \prod_{\a}<\a,\lambda>^{\c (\S)/2} \, \exp{\left(
-2 \pi^{2} \ep <\lambda, \lambda> \right) } \, , \label{int3}
\ee
for (\ref{int2}). If we shift the weight $\lambda$ by the Weyl vector
$\rho$ we obtain
\be
\sum_{\lambda}  \prod_{\a}<\a,\lambda + \rho>^{\c (\S)/2} \, \exp{\left(
-2 \pi^{2} \ep <\lambda + \rho, \lambda + \rho> \right) } \, , \label{int3a}
\ee
where it is understood that the sum is over those $\lambda + \rho$ which
are regular.

If one factors out by the action of the Weyl group, the summation over the
weight lattice can be replaced by a sum over highest weights. Then
the Weyl dimension formula relates the products appearing in the above
equations to the dimensions of the irreducible representations labelled
by the (now highest) weights $\lambda$,
\be
d(\lambda) = \prod_{\a >0}<\a , \lambda +\rho> / \prod_{\a >0} <\a , \rho >
\,\,. \label{wd}
\ee
The product $\prod_{\a >0} <\a , \rho >^{-1}$ also has an interesting geometric
interpretation as $(2\pi)^{r}$ times the Riemannian volume of the
manifold ${\bf G} / {\bf T}$. Using (\ref{wd}) we may rewrite
(\ref{int3a}) as
\be
[\prod_{\a >0}<\a, \rho>]^{\c (\Sg)} \sum_{\lambda} d(\lambda)^{\c (\S)/2}
\exp{\left(-2 \pi^{2}\ep <\lambda + \rho , \lambda + \rho> \right) } \, .
\ee
As we have not been too careful with overall factors, we drop the prefactor
$   [\prod_{\a >0}<\a, \rho>]^{\c (\Sg)}$, which is in any case a term
of the form of a standard renormalisation. The final formula for the partition
function is
\be
Z_{\Sg}(\ep) = \sum_{\lambda} d(\lambda)^{\c (\Sg)} \exp{\left(-2 \pi^{2}\ep
<\lambda + \rho , \lambda +\rho > \right) }\, . \label{int3aa}
\ee

The formula that one obtains using cutting and pasting techniques
\cite{ewym,btym,rusakov} is
rather that the partition function is given by
\be
\sum_{\lambda}  d(\lambda)^{\c (\Sg)} \,
 \exp{\left(
-2 \pi^{2}\ep C_{2}(\lambda)  \right) } \, ,
\label{int4}
\ee
where the quadratic Casimir is given by $C_{2}(\lambda) =
<\lambda + 2\rho,\lambda>$. The difference between the two calculations
of the partition function rests completely in standard renormalisations.
The area dependent renormalisation that passes one from one to the other
is $\beta = -2 \pi^{2} \ep <\rho ,\rho> = 2 \pi^{2}\ep (C_{2}(\lambda) -
<\lambda +
\rho,\lambda +\rho>)$.

\noindent \underline{Solution 2}

Consider once more (\ref{intaa}) and for simplicity fix on $SU(2)$.
Witten has given an alternative derivation of (\ref{int4}) which exhibits
the analytic structure (or lack thereof) in $\ep$ of the partition function.
Heuristically, this expression can be obtained by subtracting the singular
contribution at $\f=0$ so that we have to determine
\be
Z_{\Sg}(\ep) = \sum_{n =1}^{\infty} \int d\f \, \f^{2-2g} \exp{\left(
i\frac{ \f n}{2\pi} -
\frac{\ep \f^{2}}{16 \pi^{2}} \right)} -\int d\f \, \f^{2-2g} \exp{\left(-
\frac{\ep \f^{2}}{16 \pi^{2}} \right)} \, \d(\f) \, . \label{int2aa}
\ee
Differentiate (\ref{int2aa}) $g-1 = -\c(\Sg)/2$ times with respect to
$\ep$. This has the effect of eliminating the determinant in (\ref{int2aa}),
that is,
\bea
\frac{\partial ^{g-1} Z_{\Sg}(\ep)}{\partial \ep^{g-1}} &=&
\left(\frac{-1}{16 \pi^{2}} \right)^{g-1} \left( \sum_{n=1}^{\infty} \int d\f
\exp{\left( i\frac{\f n}{2\pi} -
\frac{\ep \f^{2}}{16 \pi^{2}} \right)} - 1 \right)\nonumber \\
& = & \left(\frac{-1}{16 \pi^{2}} \right)^{g-1}\left(-1 + \sqrt{ \frac{8
\pi^{2}}{\ep}} \sum_{n=1}^{\infty}
\exp{\left( -\frac{n^{2}}{\ep}\right)} \right) \, .
\eea

On integrating this up we obtain a formula for small $\ep$
\be
Z_{\Sg}(\ep) = \sum_{n=0}^{g-2} a_{n}\ep^{n} + a_{g-3/2} \ep^{g-3/2} + \,
{\rm exponentially \; small \; terms} \, .
\ee
The non-analyticity of this result is an indication of the singular
nature of the moduli space for $SU(2)$ \cite{ew2d}.

\noindent \underline{Non-Regular Elements of $\lt$}

In our derivation thus far we have ignored contributions to the
partition function from elements of the Lie algebra that are not
regular. Indeed it appears that these non-regular contributions are
divergent and some meaning must be given to these terms. The correct way
to handle these terms is to set them to zero.

In order to motivate this precription let us for concreteness take $\bG =
SU(2)$. Then (\ref{int2}) is
\be
\sum_{n= -\infty}^{\infty} n^{2-2g} \exp{\left( -\ep \pi^{2}
n^{2}\right)} \, , \label{reg}
\ee
and the problematic term comes from $n=0$. We have argued (and proved in
appendix B) that one obtains an $n^{2}$ from the ghost determinant and
$n^{-2g}$ from the gauge fields. If we regularise, in some way, the
gauge field determinant, while preserving all the symmetries, so that at
$n=0 $ there is no pole, then the zero from the ghost determinant will
ensure that the whole thing vanishes. At the gauge fixed level one way
to regularise is to add a small mass term for the fields $A^{\lk}$. Such
a term would still respect the left over $U(1)$ invariance and we would
obtain a thickened out version of (\ref{reg}), namely
\be
\sum_{n= -\infty}^{\infty} \frac{n^{2}}{(n + \mu)^{2g}} \exp{\left( -\ep
\pi^{2} n^{2}\right)} \, , \label{reg1}
\ee
with $\mu$ the mass. The $n=0$ term vanishes and we may take $\mu \ra
0$. We will suggest an alternative regularization (which has the same effect)
within the context of the $G/G$ model in section 3.8.

Geometrically the situation is quite clear. The Weyl group maps $\f \ra
- \f$ and consequently $n \ra -n$. Weyl group invariance is manifest in
(\ref{reg}) as only even powers of $n$ make an appearance. The worrisome
point $n=0$ is the only fixed point under the action of the Weyl group.

Likewise for $SU(n)$ the addition of a small mass to the gauge fields
will regularise in exactly the same manner as it did for the $SU(2)$
theory, as the problematic points are again those which are not acted
upon freely by the Weyl group (lie on the walls of a Weyl chamber).
Clearly, though, this is not a completely satisfactory state of affairs and
one would like to have a more conceptual understanding of what is going on
here - see section 3.8 for some further remarks.

\subsection{Symplectic Volume of $\MF(\Sg, SU(2))$ and other Observables}

One could in principle determine all the factors that we have glossed
over to give the properly normalised partition function that, when $\ep
=0$, yields the symplectic volume of $\MF$ rather than some multiple
thereof. This would, unfortunately, take us too far astray, and so we now
simply borrow all the normalisation factors that have been carefully worked
out in \cite{ewym}.

\noindent \underline{Volume of $\MF(\Sg ,SU(2))$}

The correctly normalised partition function for $\bG = SU(2)$ is
\be
2 \frac{1}{(2\pi^{2})^{g-1}} \sum_{n=1}^{\infty}
n^{2-2g} \exp{\left( -\ep \pi^{2} n^{2} \right)} \, . \label{npi}
\ee
At $\ep=0$ this is the symplectic volume of $\MF$
\bea
{\textstyle Vol}(\MF) &=& 2 \frac{1}{(2\pi^{2})^{g-1}} \sum_{n=1}^{\infty}
n^{2-2g} \nonumber \\
&= & 2 \frac{\zeta(2g-2)}{(2\pi^{2})^{g-1}} \, ,
\eea
where the Riemann zeta function is defined by
\be
\zeta(s) = \sum_{n=1}^{\infty} n^{-s} \, , \; \; \; \; {\textstyle Re}(s) >0
\, .
\ee
For $g-1 = 1,2, \dots$ one may relate the zeta function to the Bernoulli
numbers by
\be
\zeta (2g-2) = \frac{(2\pi)^{2g-2}}{2 (2g-2)!} \mid B_{2g-2} \mid \, ,
\ee
so that the volume is
\be
{\textstyle Vol}(\MF) = \frac{2^{g-1}}{(2g-2)!} \mid B_{2g-2} \mid \, .
\ee

\noindent \underline{Singularities of $\MF$ and Observables}

As we saw before $m$ differentiations of the partition function with
respect to $\ep$, evaluated at $\ep=0$, could be interpreted as the
integral over $\MF$ of an $4m$-form plus powers of the symplectic two
form required to saturate the form-degree. Differentiating (\ref{npi})
$m$ times we obtain
\be
(-1)^{m}\frac{2^{2-g}}{(\pi^{2})^{g-1-m}} \sum_{n=1}^{\infty}
n^{2-2g+2m} = (-1)^{m}\frac{2^{2-g}}{(\pi^{2})^{g-1-m}} \zeta (2g-2-2m)
\, . \label{piob}
\ee
The dimension of the
moduli space is $6g-6$ and for $m > [(3g-3)/2]$ ($[x]$ is the integer part of
$ x$) we would be
integrating a differential form of degree greater than the dimension of the
space, so for those values of $m$ one should obtain zero for the correlation
functions. The Riemann zeta function also enjoys the property that $\zeta(-2n)
 =0$ for $n \in \ZZ_{+}$. This implies that we will get a zero result for
(\ref{piob}) when $2g-2-2m<0$ or $m >g-1$. For $g \geq 2$ this
guarantees that we always obtain zero for forms whose degree is greater
than the dimension of the moduli space. However, this analysis is naive
because of the singular nature of $\MF$ (manifested by the non-analyticity
in $\epsilon$).

\subsection{Moduli Spaces with Marked Points}

The idea here is to consider
Riemann surfaces with marked points $x_{i}$. These points are ``marked''
by assigning representations $\lambda_{i}$ to these points. The way we will
do this is to place (the Fourier transform of) a co-adjoint orbit at these
points. This construction has the merit of keeping symplectic geometry to
the fore and consequently of preserving the supersymmetry (\ref{sup}).
The introduction of the marked points will mean that we are looking at
connections which are flat everywhere except at the marked points and
whose monodromy around those points is fixed. The bundles that one
obtains in this way are called parabolic and the corresponding moduli
spaces are somewhat nicer than $\MF$.

The subject of co-adjoint orbits, equivariant cohomology and fixed point
theorems is admirably addressed in \cite{bgv}. They cover many of the important
aspects of the theory that we cannot enter into here and we recommend this work
to the interested reader.

\noindent \underline{Coadjoint Orbits}

On the Lie algebra $\lg$ there is a natural conjugation action of the
group $\bG$ which extends to a natural conjugation on $\lg ^{*}$ which
is called the co-adjoint action. We identify $\lg^{*}$ with $\lg$, so that
an invariant inner product is given by the trace, $i.e.$
\be
<f,\f> \equiv Tr f \f \, , \; \; \; \; f \in \lg^{*} \, , \f \in \lg \,
,
\ee
under the action
\be
f \ra g^{-1}fg \, , \; \; \; \; \f \ra g^{-1} \f g \, , \; \; \; \forall g
\in \bG \, .
\ee

Let us fix $\lambda \in \lg^{*}$. A co-adjoint orbit through $\lambda$,
denoted $M_{\lambda}$ is the space $\bG . \lambda$, that is
\be
M_{\lambda} = \{ g^{-1} \lambda g \; ; \;  \forall g \in \bG \} \, .
\ee
If we denote the stabiliser of $\lambda$ by
$\bG (\lambda)$,
\be
\bG (\lambda) = \{ g \in \bG \, : \, g^{-1} \lambda g = \lambda \} \, ,
\ee
then the orbit $M_{\lambda}$ may be identified with the homogeneous
space $\bG / \bG (\lambda)$. We know, however, that any element of $\lg$,
and consequently of $\lg^{*}$, may be conjugated into a Cartan subalgebra
so that it suffices for us to take $\lambda \in \lt^{*}$. If $\lambda$
is regular, $\det{}_{\lk}(\ad(\lambda)) \neq 0$, then $\bG (\lambda) =
\bT$. We take $\lambda$ to be regular from now on.

\noindent \underline{Symplectic Structure and the Fourier Transform}

On the homogeneous space $\bG / \bG (\lambda)$ there is always a natural
$\bG$ invariant symplectic two-form $\Omega$ given by
\be
\Omega_{\lambda} =   \tr \, \lambda dx \wedge dx \, .
\ee
Representing the tangent vectors to $M_{\lambda}$ by elements of $\lg$
this takes on the form
\be
\Omega_{\lambda}(X,Y) = <\lambda, [X,Y] > \, , \label{symp2}
\ee
and is known as the Kirillov-Kostant form.

By the Fourier transform of an orbit one means the integral
\be
F_{M}(X) = \int_{M} \exp{\left( \frac{i}{2\pi} (<\lambda, X>\, + \,
\Omega_{\lambda}) \right)} \, . \label{ft}
\ee
In the following we will be taking expectation values of
\be
\prod_{i=1}^{n} F_{M}(\f(x_{i}))
\ee
in the topological theory for special values of $\lambda_{i}$. There is
a theorem, due to Kirillov, that states that for $\lambda =  \Lambda +
\rho $ regular and $\Lambda $ an element of the weight lattice,
\be
F_{M}(X) = j^{1/2}_{\lg}(X/2\pi) \Tr_{\lambda}(e^{X/2\pi}) \, .
\ee
We will only need $j^{1/2}_{\lg}$ evaluated on $\lt$, where it is equal
to
\be
j^{1/2}_{\lg}(X) = \prod_{\a >0} \frac{e^{<\a,X>/2} -
e^{-<\a,X>/2}}{<\a,X>} \; \;, \;\;\;\; \; X \in \lt \, .
\ee
The function $j_{\lg}$ is the Jacobian that passes one from the Lie
algabra to the Lie group (and vice-versa). Indeed the denominator we
have seen before, it is $\det{}_{\lk}(\ad(X))$. The numerator will also
play an important role in our study of the $G/G$ models.

\noindent \underline{A Supersymmetry}

If we introduce Grassman variables to represent the differential forms
on $M_{\lambda}$ we may express the action in (\ref{ft}) for each orbit as
\be
<\lambda , \f(x) > + < \lambda, \trac{1}{2}[\ga, \ga ] > \, . \label{coact}
\ee
This action enjoys the following supersymmetry
\be
\d \lambda = [\lambda ,\ga] \, , \; \; \; \d \ga = \f(x) + \ga^{2} \, \;
\; \; \d^{2} = {\cal L}_{\f(x)} \, . \label{sup2}
\ee
Notice that, just as for the supersymmetry (\ref{sup}), this supersymmetry
closes on gauge transformations (conjugation), parameterised by $\f$ but
now $\f$ is evaluated at the marked point $x$. Again one is working
equivariantly and the action of the Fourier transform is known to be the
integral of an equivariantly closed two-form. The action for the
topological theory, as we saw, represented a generator for the second
cohomology group of the moduli space of flat connections $H^{2}(\MF)$.
One consequence of the addition of the co-adjoint orbit is that there is
another generator of the second cohomology group of the new moduli space
to take into account. The new generator is the co-adjoint action itself.

\noindent \underline{$k{\cal A}\times \prod_{i=1}^{s} \a_{i}M_{\lambda_{i}}$}

Combining all the spaces we see that we are now working on the space of
all connections ${\cal A}$ together with the co-adjoint orbits
$M_{\lambda_{i}}$ at the marked points $x_{i}$. Denote this space, with
its natural symplectic two-form given by (\ref{symp}) and (\ref{symp2}),
by ${\cal A} \times \prod_{i=1}^{s}M_{\lambda_{i}}$. We generalise the
situation a little by considering the same spaces but with $k$ times the
symplectic form (\ref{symp}) and with $\a_{i}$ times the symplectic
two-form of each of the orbits $M_{\lambda_{i}}$. $k{\cal A}\times
\prod_{i=1}^{s} \a_{i}M_{\lambda_{i}}$ denotes the space with the new
symplectic two-form.

For the moment we do not insert the co-adjoint orbits into the path
integral but ask what result we should expect for the volume of $\MF$ on
taking $k$ times (\ref{symp})? This is easily answered, a glance at
(\ref{sympvol}) shows us that we should get $k^{\dim \MF /2}$ times the
original volume. In the case of $\bG = SU(2)$ this means that we should
obtain
\be
{\textstyle Vol}_{k}(\MF) = k^{3g-3} 2 \frac{1}{(2\pi^{2})^{g-1}}
\sum_{n=1}^{\infty} n^{2-2g} \, .
\ee
So as not to introduce $k$ dependence into the supersymmetry
transformation rules (\ref{sup}) one should use $k$ times the original
topological action as the new action or, put another way, $k$ times the
generator of $H^{2}\left(\MF (\Sg , \bG ) \right)$.

Let us see, for a co-adjoint orbit, what taking $\a$ times the
generator of the second cohomology group yields. Firstly, the
generator in question is not just the symplectic two form (\ref{symp2})
but is its invariant extension, the action (\ref{coact}). So we really
want to consider $\a$ times this action. We can use the original
symplectic two-form as long as we remember to multiply the Fourier
transform by $\a^{m}$, where the dimension of the orbit is $2m$. The
remaining integral is the Fourier transform
of $\a X$ so we have
\bea
\int_{M} \exp{\left( \frac{i\a}{2\pi}(<\lambda, X>\, + \,
\Omega_{\lambda}) \right)} &=& \a^{m} \int_{M} \exp{\left(
\frac{i}{2\pi}(\a< \lambda, X>\, + \,  \Omega_{\lambda}) \right)} \nonumber \\
&=& \a^{m} \Tr_{\lambda}\left( e^{\a X/2\pi} \right) j_{\lg}^{1/2}(\a
X/2\pi) \, .
\eea

\noindent \underline{Some Formulae for $SU(2)$}

It is straightforward to push through, formally, for arbitrary compact
groups, all of the calculations we are about to make for $SU(2)$. There
is one aspect that arises in the more general situation, that we will not
see here, namely that there is more than one preferred symplectic two
form on the co-adjoint orbit (indeed there are a ranks worth). However,
we will stick with the simplest
example and for the rest of this section $\bG = SU(2)$.

We wish to evaluate the following path integral
\bea
 & & <\prod_{i=1}^{s} F_{M_{\lambda_{i}}}(\f(x_{i}))> \nonumber \\
 & & = \; \; \int_{\Phi} \, \exp{\left( \frac{ik}{4\pi^{2}}\int_{\Sg} \Tr
\left(  i \f F_{A} + \frac{1}{2} \p
\wedge \p \right)\right)} \,  \prod_{i=1}^{s}
F_{M_{\lambda_{i}}}(\f(x_{i})) \, .
\eea
Following the manipulations that allowed us to evaluate the Yang-Mills
path integral leads to
\be
<\prod_{i=1}^{s} F_{M_{\lambda_{i}}}(\f(x_{i}))> = k^{3g-3} 2
\frac{1}{(2\pi^{2})^{g-1}} \sum_{n=1}^{\infty} n^{2-2g} \, \prod_{i=1}^{s}
F_{M_{\lambda_{i}}}(\frac{4\pi^{2} n}{k}) \, ,
\ee
where the position dependence has dropped out, as one would expect on
general grounds, and, as it is $k\f$ that appears in the action, the
localisation is $k\f = 4\pi^{2} n$.

For $SU(2)$ we may label the unitary irreducible representations by
their dimension $j$ and we have the following simple formulae,
\be
\Tr_{j}\left( e^{\frac{2 \pi i n}{k}} \right) = \frac{\sin{\pi j
n/k}}{\sin{\pi n/k}} \, ,
\ee
and
\be
j^{1/2}_{\lg}(2\pi n/k) = \frac{\sin{\pi n/k}}{\pi n/k} \, ,
\ee
Making use of these equations we find
\be
<\prod_{i=1}^{s} F_{M_{\lambda_{i}}}(\f(x_{i}))> =
k^{3g-3 +s} 2
\frac{1}{(2^{g-1} \pi^{2g-2+s})} \sum_{n=1}^{\infty} n^{2-2g-s}
\prod_{i=1}^{s} \sin{\pi j_{i} n/k } \, . \label{sympob}
\ee
This is a formula for the volume of the moduli space of connections
on a Riemann surface $\S$, flat on $\S\setminus\{x_{k}\}$ and with
prescribed holonomies around the
$s$ points $x_{k}$ corresponding to the representations $j_{k}$
(or, in other words, a moduli space of parabolic bundles on $\S$).
Let us denote this moduli space by ${\cal
M}_{SU(2)}(s)$ (though we should also indicate the $s$ representations
as well). The dimension of this space is the same as the dimension of
the space of flat connections ${\cal M}_{SU(2)}$ plus the sum of the
dimensions of the co-adjoint orbits. As we are restricting ourselves to
regular elements, the co-adjoint orbits are simply $SU(2)/U(1) \equiv
S^{2}$. We have therefore
\be
\dim {\cal M}_{SU(2)}(s) = \dim {\cal M}_{SU(2)} \, + 2s \, .
\ee
One sees that the power of $k$ in (\ref{sympob}) is simply $\dim {\cal
M}_{SU(2)}(s)/2 $ as it should be.

The formula (\ref{sympob}) has been obtained by Witten using an approach
dual to that employed here. Witten considers Riemann surfaces with
boundary and on the boundary places conjugacy classes $\Theta$ of $\bG$
of order $k$. The duality arises on thinking of the Riemann surface with
boundary as arising from cutting out discs centred at the fixed points
of a closed Riemann surface. The conjugacy classes are then identified
with the orbits.

In order to obtain expectation values including the new generator
of the second cohomology group $H^{2}(\M_{SU(2)}(s))$, one should
differentiate the partition function with respect to the $\a_{i}$
and then set $\a_{i}=1$ at the end. For example, with one marked point
and differentiating once, yields (again for $SU(2)$)
\be
k^{3g-2}\frac{1}{2^{g-1}\pi^{2g-2}}\sum_{n=1}^{\infty} n^{2-2g} \cos \pi jn/k
\ee
as the intersection number of $\M_{SU(2)}(1)$ involving one insertion of
$\Omega_{\lambda}$ and $(3g-3)$ insertions of the symplectic form (\ref{symp})
on the space of gauge fields.

\section{The ${\bf G/G}$ Model and the Verlinde Formula}

In the previous sections we have dealt with integrals over Lie algebras
and path integrals of quantum field theories involving Lie algebra valued
fields. There are, however, certain interesting classes of field theories
involving group valued fields, most notably Wess-Zumino-Witten (WZW)
models and their relatives.
For those theories,
the integration formulae of the previous sections need to be modified
appropriately and the aim of this section is to establish
the Weyl integral formula for Lie groups and to use it to calculate
the partition function of the so-called $G/G$ model,
a non-linear counterpart of the theories (BF and Yang-Mills)
studied in the previous section. The main motivation for studying this
particular model stems from topological field theory, in particular
from its relation with Chern-Simons theory in three dimensions and the
Verlinde formula for the number of conformal blocks of a rational
conformal field theory. For more on this part of the story see \cite{ewwzw}
and \cite{btver}. Here we will primarily be interested in the $G/G$ model
as a rather non-trivial but nevertheless exactly solvable model
(via the method of `Abelianization') in its own right.

We begin by reviewing some basic facts about the WZW model and gauged
WZW models in general. We then specialize to the $G/G$ model and discuss
some of its properties (like its topological nature). We also explain
briefly the relation among the $G/G$ model, Chern-Simons theory, and
the Verlinde formula. In the following we then set out
to calculate the partition function
in a two-step procedure, by first reducing it to an Abelian theory via
a suitable choice of gauge and then evaluating the resulting (simple)
Abelian theory to obtain the Verlinde formula.

\subsection{The WZW model}

The WZW model is one of the most important examples of a (rational)
conformal field theory. The fields are maps $g: \S\ra\bG$
from a two-dimensional
manifold to a compact Lie group (which we take, as above, to be $SU(n)$
or at least simple, connected, and simply connected). If $\S$ has no
boundary, the action $kS_{G}(g)$ of the WZW model (at level $k\in\ZZ$)
is the sum of two terms,
\be
kS_{G}(g) = kS_{0}(g)-ik\Gamma(g)\;\;.\label{2}\\
\ee
The first of these is the standard non-linear sigma model action,
\be
S_{0}(g) = -\trac{1}{4\pi}\int_{\S}d^{2}\! z \g\dz g \g\dzb g \;\;,
\label{3}
\ee
written here in terms of complex coordinates on $\S$. From now on
a trace is understood to be implicit in integrals of Lie algebra
valued fields. Apart from that we follow the
conventions of \cite{ewwzw}.

The second is the so-called Wess-Zumino (WZ)
term. It was introduced originally in four dimensions to
incorporate the effects of chiral anomalies (like $\pi^{0}\ra 2\gamma$)
in low-energy efective Lagrangians \cite{wz}.
Its two-dimensional counterpart appears in
the `non-Abelian bosonization' of fermions and its topological
significance was realized and explained in \cite{ewwz}. The WZ term
can usually only be written locally as the integral of a $2d$ density
(although its variation is always local) and it is convenient to
introduce a three-manifold $N$ bounded by $\S$, to extend the
field $g$ in some way to a map $g_{N}$ from $N$ to $\bG$, and to then write
the WZ term as
\be
\Gamma(g)= \trac{1}{12\pi}\int_{N}d^{3}\!x \epsilon^{ijk}
             \g_{N}\del_{i}g_{N} \g_{N}\del_{j}g_{N} \g_{N}\del_{k}g_{N}
\label{4}\;\;.
\ee
That this is essentially independent of the choice of extension or the choice
of $N$ follows from the fact that the integral of the above three-form over
a closed three-manifold $M$ is a topological term measuring the winding
number of the map $g: M\ra\bG$ and taking values in $2\pi\ZZ$ ($\pi_{3}
(\bG) = \ZZ$). Hence (for integer $k$) the amplitude $\exp -kS_{G}$ is
independent of these choices. We will therefore mostly denote $g_{N}$
simply by $g$ - a notational convenience which has its pitfalls as we will
see in section 3.6.

While the first term all by itself is conformally invariant classically,
it is the presence of the second term (with precisely the coefficient given
in (\ref{2})) that ensures the conformal invariance of the quantum theory.
In fact, we will see below that for this coefficient the action has two
commuting Kac-Moody symmetries. These survive the quantization and
ensure conformal invariance via the Sugawara construction.

The most important property of the WZW action is its behaviour under
pointwise multiplication of the maps $g$, given by the so-called
Polyakov-Wiegmann (PW) formula (\cite{pw})
\bea
S_{G}(g_{1}g_{2})&=&S_{G}(g_{1})+ S_{G}(g_{2}) + C(g_{1},g_{2})
\bmod 2\pi i\ZZ \;\;,\label{pw1}\\
C(g_{1},g_{2})   &=&
\trac{1}{2\pi}\int_{\S} d^{2}\! z\, \g_{1}\dz g_{1} g_{2}\dzb \g_{2}
\;\;.\label{pw2}
\eea

For example, by taking $g_{2}$ `small', the PW formula implies immediately
the equations of motion
\be
\d S_{G} = 0 \Rightarrow
  \dzb (\g\dz g) = 0 \leftrightarrow  \dz (\dzb g\,\g) =0 \;\;.\label{wz1}
\ee
Likewise, because only the $\del$ derivative of $g_{1}$ enters into
(\ref{pw2}) and only the $\dbar$ derivative of $g_{2}$, (\ref{pw1})
shows that the WZW action is invariant under the right multiplication of
$g(z,\bar{z})$ by a locally holomorphic $\bG_{\CC}$ valued map $g_{2}(z)$,
and under left  multiplication by $g_{1}(\bar{z})$. Thus, altogether
we have a $\bG_{L}(\bar{z})\times\bG_{R}(z)$ invariance which gets promoted
to two commuting Kac-Moody symmetries at the quantum level. The equations
of motion (\ref{wz1}) can be regarded as the corresponding conservation laws
for the currents
\be
j_{z}=\g\dz g\;\;,\;\;\;\;\;\;j_{\bar{z}} =\dzb g\,\g
\ee
generating right and left translations respectively.

\subsection{The gauged WZW model}

Usually, when one has a Lagrangian with a global symmetry, it is possible
to find an extension of the original Lagrangian containing additional fields
like gauge fields which is invariant under local symmetry transformations.
In the simplest cases this just amounts to replacing ordinary
derivatives by covariant
derivatives (minimal coupling).

In the case of the WZW model this procedure indeed works for the first term
of the action for any subgroup of $\bG_{L}\times\bG_{R}$. The WZ term, however,
having its origin in an anomaly, can only be gauged for those subgroups $\bF$
of $\bG_{L}\times\bG_{R}$ which are `anomaly free', i.e. which satisfy
\be
\tr_{L}f_{1}f_{2} = \tr_{R}f_{1}f_{2} \label{wz2}
\ee
for all $f_{1},f_{2}\in\lf$, the Lie algebra of $\bF$, $\tr$ denoting the
trace in the adjoint representation. This condition is fairly restrictive but
can be satisfied by choosing $\bF$ to be a diagonal subgroup of $\bG\times\bG$
which we think of as a subgroup $\bH\ss\bG$ acting via $g\ra hgh^{-1}$.
It can also be satisfied
by choosing $\bF$ such that both sides of (\ref{wz2}) are separately zero
(e.g. by taking upper or lower triangular matrices), but
it is the former possibility that has attracted more attention and that
we are interested in. The relation of (\ref{wz2}) with
equivariant cohomology and the `optimal' gauging of anomalous subgroups
are discussed in detail in \cite{ewwzw} and we will not repeat this here.

Introducing an ${\lh}$ valued gauge field $A$ (where we have split the
Lie algebra as $\lg = \lh \oplus \lk$),
the action $S_{G/H}$ of the gauged WZW model is
\bea
S_{G/H}(g,A) &=& S_{G}(g) + S_{/H}(g,A)\label{5}\;\;,\\
S_{/H}(g,A) &=& -\trac{1}{2\pi}\int_{\S}d^{2}\! z
(\az\dzb g\g -\azb\g\dz g + \az\azb - \g\az g\azb)\;\;.\nonumber
\label{6}
\eea
This action has been shown \cite{gk} to provide a field-theoretic realization
of the GKO coset model construction \cite{gko}. In particular, the choice
$\bG= SU(2),\bH=U(1)$ is related to parafermions, while the discrete series
of Virasoro unitary minimal models is obtained by choosing $\bG= SU(2)_{k}
\times SU(2)_{1}$ and $\bH = SU(2)_{k+1}$ (the subscript referring to the
level).

By the variation of the gauge field, the $\lh$ components of the
covariantized currents  ($d_{A} = D_{z}dz + D_{\bar{z}}d\bar{z}$)
\be
J_{z}=\g D_{z} g\;\;,\;\;\;\;\;\;J_{\bar{z}} = D_{\bar{z}} g\,\g
\ee
are set to zero as equations of motion (or constraints),
\be
J_{z}^{\lh} = J_{\bar{z}}^{\lh} = 0\;\;,
\ee
as behoves the currents corresponding to a local gauge symmetry. Using
these constraints, the $\lh$ and $\lk$ parts of the remaining $g$
equations
of motion can be separated and read
\be
F_{A}=0\;\;,\;\;\;\;\;\;D_{\bar{z}}J_{z}^{\lk} = D_{z}J_{\bar{z}}^{\lk}=0
\;\;.
\ee
Finally, we note that in both the gauged and the ungauged WZW model the
energy-momentum tensor is known to be of the Sugawara - Sommerfield form,
i.e.~quadratic in the currents.

\subsection{The $\bf G/G$ model}

Taking ${\bf H}=\bG$ in (\ref{5}) one obtains the action of the
$G/G$ model. For later purposes it will be convenient to have this
action written in terms of differential forms,
\bea
S_{G/G}(g,A)&=&-\trac{1}{8\pi}\int_{\S}\tr \g d_{A}g *\g d_{A}g -i\Gamma(g,A)
\label{diff}\\
\Gamma(g,A)&=& \trac{1}{12\pi}\int_{N}\tr (\g dg)^{3}-\trac{1}{4\pi}\int_{\S}
\tr\left(A(dg\,\g + \g dg) + A\g A g\right)\;\;.\nonumber
\eea
The action is invariant under the transformations
\be
g\ra g^{h}\equiv h^{-1}g h \;\;,\;\;\;\;\;\;A\ra A^{h} = h^{-1}Ah + h^{-1}dh
\;\;,\label{wz3}
\ee
where $h = h(z,\zb)$.

In spite
of what the name may suggest, this model is not completely empty. However,
it is also not a full-fledged quantum field theory but a theory with only
a finite number of quantum mechanical degrees of freedom
(this can easily be checked by counting degrees of freedom and
constraints).
This model has been discussed in relation with Chern-Simons
theory by Verlinde and Verlinde \cite{verver} some years ago and has
recently attracted renewed attention as a rather interesting and
rich source of topological field theories, see e.g.~\cite{sp,ewwzw,btver}.

A priori, the topological nature of this model is far from obvious,
however, as the action $S_{G/G}$ is neither metric independent nor of the
form expected of a cohomological field theory. Nevertheless, the metric
independence of the partition function and certain correlators can
(formally) be established by direct calculation \cite{ewwzw} and we
will sketch the argument below.

\nul{Equations of Motion}

Further circumstantial evidence for the topological
nature of the $G/G$ model can be obtained from an examination of the
equations of motion. These can be written as
\be
F_{A} = 0\;\;,\;\;\;\;\;\;d_{A}g = 0 \leftrightarrow A^{g}=A\;\;,
\ee
i.e.~classical configurations are gauge equivalence classes of pairs
$(A,g)$ where $A$ is flat and $g$ is a symmetry of $A$. As in
genus $>1$ generically flat connections are irreducible, the classical
phase space is the moduli space of flat connections with fibers
corresponding to the isotropy groups attached at the reducible points.
This is very reminiscent of the phase space of Chern-Simons theory on
a three-manifold of the form $\S\times\RR$ and even more of that
of two-dimensional non-Abelian BF theory discussed in the first part
of these lectures. Its action
\be
S_{BF} = \int_{\S}\tr BF_{A}
\ee
(what we call $B$ here was called $\f$ before - henceforth $B$ will
refer to an ordinary Lie algebra valued scalar while $\f$, which will
appear shortly, denotes a compact, torus-valued scalar)
implies the equations of motion
\be
F_{A} = 0\;\;,\;\;\;\;\;\;d_{A}B = 0 \leftrightarrow \d_{B}A = 0\;\;.
\ee
As the $A$ equations of motion in the two models are identical and
the $B$ equation of motion is precisely the infinitesimal version of the
equation of motion $A^{g}=A$ of the $G/G$ model, this suggests that
the $G/G$ model is some (compact or non-linear) deformation of
BF theory. That this is indeed the case (the precise statement being
that in the $k\ra\infty$ limit the $G/G$ model reduces to the BF model)
can be established directly. Alternatively, it
follows from the equivalence of Chern-Simons theory on $\SS$ with
the $G/G$ model on $\S$ established recently in \cite{btver}.
We will come back to the relation among these theories below.

\nul{Metric Independence}

We will now examine the question of metric dependence of the partition
function (the litmus test for a topological field theory). When varying
an action with respect to the metric $g_{\alpha\beta}$ one obtains the
energy momentum tensor $T_{\alpha\beta}$ of the theory via
\be
\d S = \trac{1}{2}\int \sqrt{g} \d g^{\alpha\beta} T_{\alpha\beta}\;\;.
\ee
Doing this in the case of the $G/G$ model one finds (confirming the
Sugawara form)
\be
\d S_{G/G} = \trac{1}{8\pi}\int_{\S}\!d^{2}\!z\,\left(
\d g_{\zb\zb} g^{\zb z} \tr J_{z}J_{z} + \d g_{zz} g^{z\zb}\tr J_{\zb}J_{\zb}
\right)\;\;.
\ee
Thus the variation of the partition function with respect to the metric
will lead to an insertion of this expression into the path integral. Noting
that a variation of $\exp -k S_{G/G}(g,A)$ with respect to, say, $\azb$
will lead to an insertion of $J_{z}$, one sees that the metric variation
of the partiton function (the vacuum expectation value of the energy momentum
tensor) can be written as a total derivative in the space of gauge fields
and is hence (formally) zero,
\bea
\d Z_{\S}(S_{G/G}) &=& \int\!DA\,Dg (-k\d S_{G/G}(g,A)) \ex{-kS_{G/G}(g,A)}
\nonumber\\
&=& \trac{1}{4}\int\!DA\,Dg \frac{\d}{\d\azb^{a}}
\left(\int_{\S} \d g_{\zb\zb} g^{\zb z} J_{z}^{a}\ex{-kS_{G/G}(g,A)}\right)
\nonumber\\
&-& \trac{1}{4}\int\!DA\,Dg \frac{\d}{\d\az^{a}}
\left(\int_{\S} \d g_{zz} g^{z\zb} J_{\zb}^{a}\ex{-kS_{G/G}(g,A)}\right)
\nonumber\\
&=& 0\;\;.
\eea
By the same argument, correlation functions of traces of $g(z,\zb)$
are also metric independent (and, in fact, they are known to reproduce the
fusion rules of the WZW model \cite{btver}). It is clear from the above,
however, that correlation functions of operators involving Wilson loops
of the gauge field $A$
(which a priori are perfectly respectable gauge invariant and metric
independent
`topological' observables) will not necessarily be metric independent
but rather depend on the length of the loop. This is an interesting
phenomenon which has no counterpart in either of the more standard types
of topological field theories.

\nul{Supersymmetric Extension}

As we have mentioned above, it is appropriate to think of the $G/G$ model
as a non-linear deformation of BF theory. In fact, a comparison of the
equations of motion suggests that BF theory is a tangent space
approximation to the non-linear $G/G$ model.
This raises the question whether
in this model there exists a counterpart of the supersymmetry which is
present in
BF and Yang-Mills theory, c.f.~(2.13)-(2.15) and involves the
symplectic form $\int_{\S}\p\p$ on $\cal A$. This is indeed the case.
As the variation of the action $S_{G/G}$ with respect to the gauge fields
is
\be
\d S_{G/G}(g,A) = \trac{1}{2\pi}\int_{\S} \tr (J_{z} \d\azb -
J_{\zb}\d\az)\;\;,
\ee
the combined action
\be
S(g,A,\p) = S_{G/G}(g,A) - \trac{1}{2\pi}\int_{\S} \p_{z}\p_{\zb}
\ee
is invariant under the transformations
\bea
&&\d\az = \p_{z}\;\;,\;\;\;\;\;\;\d\p_{z} = J_{z} \equiv \g D_{z}g\;\;,
\nonumber\\
&&\d\azb= \p_{\zb}\;\;,\;\;\;\;\;\;\d\p_{\zb} = J_{\zb}\equiv D_{\zb}g\,\g\;\;,
\label{ggsusy1}
\eea
supplemented by $\d g =0$. What is interesting about this supersymmetry is
that it does not square to infinitesimal gauge transformations, like its
Yang-Mills counterpart, but rather to `large' gauge transformations,
\bea
&&\d^{2}\az = \az^{g} - \az\;\;,\;\;\;\;\;\;\d^{2}\p_{z} = \g\p_{z} g - \p_{z}
\;\;,\nonumber\\
&&\d^{2}\azb = \azb - \azb^{\g}\;\;,\;\;\;\;\;\;\d^{2}\p_{\zb} =
\p_{\zb} - g\p_{\zb} \g\;\;.
\label{ggsusy2}
\eea
This suggests that some global counterpart of (the infinitesimal)
equivariant cohomology could provide the right interpretational
framework for this model - an issue that appears to merit further
investigation.

Alternatively, one can put the supersymmetry into slightly more familiar form
by noting that the supersymmetry operator $\d$ can be written as the sum
of two nilpotent operators $Q$ and $\bar{Q}$,
\be
\d = Q + \bar{Q}\;\;,\;\;\;\;\;\; Q^{2} = \bar{Q}^{2} = 0\;\;,
\ee
where e.g.
\bea
QA_{z} = \p_{z}\;\;&,&\;\;\;\;\;\;Q \azb =0 \;\;, \nonumber\\
Q\p_{z} = 0 \;\;&,&\;\;\;\;\;\; Q\p_{\zb} = J_{\zb}\;\;.
\eea
It can be shown that both this Dolbeault-like symmetry and the
$G/G$ action are infinite dimensional counterparts of the theory of
equivariant Bott-Chern currents discussed e.g.~by Bismut in \cite{bismut}.
Note also, that both (\ref{ggsusy1}) and (\ref{ggsusy2})
exhibit the chiral nature of the supersymmetry (and hence that of the
$G/G$ model).

\subsection{Relation with Chern-Simons Theory}

While we have seen above that the $G/G$ model has topological correlation
functions, it would be nice to know {\em a priori} what topological
quantity these correlation functions calculate. It turns out that, with
proper normalizations, these topological invariants are integers - the
dimensions of certain vector spaces which can be associated to the data
$(\S,\bG,k)$. From the conformal field theory point of view (which we shall
not pursue here, see e.g. \cite{ewwzw}), these vector spaces are the
spaces of conformal blocks of the $\bG$-WZW model on $\S$ at level $k$,
and a general formula for their dimension has been derived by E.~Verlinde
\cite{ver}. We will recall the Verlinde formula below.

On the other hand, these vector spaces also arise as the
Hilbert spaces of a three-dimensional topological gauge theory, Chern-Simons
theory, when canonically quantized on three-manifolds of the form
$\S\times\RR$.\footnote{The discovery that the Hilbert spaces of
Chern-Simons theory are the conformal blocks of the WZW model (`satisfy the
axioms of a modular functor') was the starting point for the interest
in Chern-Simons theory as a topological field theory \cite{ewcs}.}
We thus need to understand (a) what is the relation between Chern-Simons
theory and the $G/G$ model and (b) why, as a consequence, the correlators
of the $G/G$ model calculate the dimensions of the Chern-Simons Hilbert
spaces.
We will now sketch the answer to these questions. This discussion
is, however, not meant to be self-contained (the emphasis in these
lectures being on the techniques to deal with the $G/G$ model itself)
- see \cite{btver} for details and e.g.~\cite{ewcs,emss,axcs,pr} for
background information.

Choosing a closed oriented three-manifold $M$ and a compact gauge
group $\bG$ (which we will assume to be simply-connected so that
any principal $\bG$-bundle over $M$ is trivial), the Chern-Simons
action for $\bG$ gauge fields $\unA$ on $M$ (we reserve the notation $A$
for spatial gauge fields) is defined by
\be
kS_{CS}(\unA) =\trac{k}{4\pi}\int_{M}\tr(\unA d\unA + \trac{2}{3}\unA^{3})\;\;.
\ee
The trace (invariant form on the Lie algebra $\lg$ of $\bG$)
is normalized in such a way that invariance of $\exp ik S_{CS}$ under
large gauge transformations requires $k\in\ZZ$.

The action $S_{CS}$
is a non-trivial metric-independent classical action which gives rise to
one of the richest topological quantum field theories.
For example, the partition function $Z_{M}(S_{CS})$ is a topological
invariant of the three-manifold $M$ which can be evaluated both perturbatively
and non-perturbatively and has been studied intensely, see \cite{nncs}
for recent work in this direction.
Moreover, correlation functions of Wilson loops are invariant under
deformations (isotopies) of these loops and hence give rise to
generalized knot and link invariants (generalized, because one is not
necessarily restricted to knots in $S^{3}$).

\nul{Chern-Simons Theory on $\SS$}

Of interest to us here are the cases where the three-manifold is either
of the type $\SS$ or of the type $\S\times\RR$ for some closed
two-dimensional surface $\S$. In the former case, \CS\ theory
is actually equivalent to the $G/G$ model on $\S$ (answering question
(a) above). That some such equivalence should hold can be seen as
follows. First of all, it is convenient to regard the path integral
on $\SS$ as the trace of an amplitude on $\SI$, with some boundary
conditions on the spatial (i.e.~$\az, \azb$) components of the
gauge fields at $\Sa$ and $\Sb$. Denoting
by $A_{0}$ the component of $\unA$ along the $S^{1}$, we can thus think
of the \CS\  path integral as being given by an action which is a functional
of $A_{0}$ and `time' independent fields $\az$ and $\azb$.
Moreover,
the only gauge invariant degree of freedom carried by $A_{0}$ is (the
conjugacy class of) its holonomy, the path-ordered exponential
\be
g(A_{0}) := P\ex{\oint_{S^{1}}A_{0}}\;\;.
\ee
For each $A_{0}$ this is a map $g(z,\zb)$ from $\S$ to $\bG$. Because
of the gauge invariance of \CS\ theory, it should thus in principle be possible
to trade $A_{0}$ for $g$ as a fundamental field. In that way one arrives
at a topological action from which all the $S^{1}$ dependence has disappeared
and
which is a functional $S(g,A)$ of a $\bG$-valued field $g$ and a $\lg$-valued
connection $A$ on $\S$. At this point it should be quite plausible that this
is nothing but the action $S_{G/G}(g,A)$ of the $G/G$ model, and this can
indeed
be confirmed by explicit calculation \cite{btver}.

\nul{Canonical Quantization of Chern-Simons Theory}

To address question (b) we consider the second case,
$M=\S\times\RR$. On such manifolds, \CS\ theory can be
subjected to a canonical analysis. Upon choosing the gauge $A_{0}=0$,
one determines the classical reduced phase space to be the moduli space
$\MF$ of flat connections on $\S$. As we saw before,
this is a compact symplectic space
(unlike the non-compact cotangent bundles one usually obtains as phase
spaces in classical mechanics)
which becomes K\"ahler once one chooses a complex structure on $\S$.
In order to quantize this system, one can appeal to the techniques of
geometric quantization which have been devised to deal
with precisely such situations - see \cite{axcs} for a discussion of
geometric quantization in the present context.
According to geometric quantization,
the Hilbert space will be the space of holomorphic sections of a
line bundle over $\M$ whose curvature is ($i$ times) the symplectic form
of $\MF$.\footnote{It follows from Quillen's calculation \cite{qdet}
and the fact that
the symplectic form for `level $k$' \CS\ theory is $k$ times the fundamental
symplectic form $\frac{1}{2\pi}\int_{\S}\d A \d A$, that the line bundle
in question is (for $SU(n)$) the $k$-th power of the determinant line bundle
associated to the family of operators $\{\bar{\del}_{A}\}$.}
In \cite{ewcs} and \cite{verver} it is shown that
the space $V_{g,k}$ of holomorphic sections of this line bundle
(the Hilbert space of
\CS\ theory) coincides with the space $V_{g,k}$
of holomorphic blocks of the $\bG_{k}$ WZW model.

What is important for us is the fact that the dimension of this vector
space is given by a path integral of \CS\ theory. In fact, since the
Hamiltonian of \CS\ theory is zero (like that of any generally covariant
or topological theory), the statistical mechanics formula
\be
Z_{\SS} = \Tr e^{-\beta H}
\ee
for a circle of radius (imaginary time) $\beta$ reduces to
\be
Z_{\S\times S^{1}}(S_{CS},k) = \dim V_{g,k}\;\;.
\ee
Combined with the above-mentioned equivalence of \CS\ theory on
$\SS$ with the $G/G$ model on $\S$, this implies that the
partition function of the $G/G$ model indeed calculates the dimension
of the Chern-Simons Hilbert space, answering (b).

\nul{Marked Points and the Verlinde Formula}

For the topological correlation functions of the traces of $g(z,\zb)$,
the story is quite similar. From what we said above it is clear that
the gauge invariant operator $\tr_{R} g$ ($R$ a representation of $\bG$)
of the $G/G$ model corresponds
to the trace of a `vertical'
Wilson loop on $\SS$ in \CS\ theory. Hence \cite{ewcs} a correlator of $s$
such operators in the $G/G$ model can be thought of as the dimension
of a vector space $V_{g,s,k}(R_{1},\ldots,R_{s})$ associated to a
genus $g$ surface with $s$ marked points labelled by the representations
$R_{i}$. Again this is an integer and, in particular, a topological
invariant associated to these data.

{}From conformal field theory, formulae for these dimensions are known
(see \cite{ver,ms}). For instance, in the case $\bG=SU(2)$ one has
(labelling the $(l+1)$-dimensional representation of $SU(2)$ by
$l\in\ZZ_{+}$)
\be
\dim V_{g,s,k}(l_{1},\ldots,l_{s}) =
(\trac{k+2}{2})^{g-1}
\sum_{j=0}^{k}\left(\sin\trac{(j+1)\pi}{k+2}\right)^{2-2g-s}\prod_{i=1}^{s}
\sin\trac{(j+1)(l_{i}+1)\pi}{k+2}\;\;.\label{ver2}
\ee
This expression has several notable
(and non-obvious) features, not the least of which is that it is indeed
an integer. Its generalization to $\bG = SU(n)$
(and no arked points, for simplicity) is
\be
\dim V_{g,k} = \left(n(k+n)^{n-1}\right)^{g-1}\sum_{\la\in\Lambda_{k}}
\prod_{\a\in\Delta}(1-e^{i\frac{\a(\la + \rho)}{k+n}})^{1-g}\;\;,\label{i3}
\ee
where $\Lambda_{k}$ denotes the space of integrable highest weights
at level $k$, $\Delta$ the set of roots of $\bG$, and $\rho$ the Weyl vector
(half the sum of the positive roots).  It is these somewhat daunting formulae
which we will be able to derive (up to the standard renormalization
factors $\sim ({\rm something})^{g-1}$)
rather straightforwardly by evaluating
the $G/G$ partition function via the Weyl integral formula. In principle,
the normalization can also be fixed without invoking conformal field
theory or the representation theory of loop groups, but we will not
attempt this here.

\subsection{The Weyl Integral Formula for Lie Groups}

The large gauge invariance of the $G/G$ model allows one to solve
this model completely via the method of Abelianization. For this
we will need the Lie group analogue of the integral formula used
in section 2 to solve Yang-Mills theory.

\nul{Some Lie Group Theory}

To write down and explain the \wif\ we will have to introduce some
more notation.
Thus let $\bG$ be a compact Lie group (which we will take to be $SU(n)$ for
concreteness) and $\bT$ a maximal torus of $\bG$, i.e. a
maximal Abelian subgroup. For $\bG = SU(n)$, $\bT\sim U(1)^{n-1}$,
which can be realized by diagonal matrices in the fundamental
representation of $SU(n)$. Its Lie algebra is the Cartan subalgebra $\lt$
which played a prominent role in the previous sections.

Now, as is well known, any unitary matrix can be diagonalized. More
abstractly, one says that any element of $\bG$ can be conjugated into $\bT$.
The residual conjugation
action of $\bG$ on $\bT$ (conjugation by elements of $\bG$ which leave $\bT$
invariant) is that of a finite group, the Weyl group
$W$, that we encountered in section 2.3. From the above description it follows
that the Weyl group can be
thought of as the quotient $W=N(\bT)/\bT$, where
$N(\bT)= \{g\in\bG: \g t g\in \bT \;\forall t \in \bT\}$
denotes the normalizer of $\bT$ in $\bG$) and the quotient by $\bT$ is to be
taken because the conjugation action of $\bT$ on itself is trivial.
In the case of $SU(n)$, $W$ is the permutation group $S_{n}$ on $n$ objects
acting on an element of $\bT$ by permutation of the diagonal entries.

Thus, if we are given a conjugation invariant real or complex valued function
on $\bG$ (a class function), then it is determined entirely by its
restriction to $\bT$ (where it is $W$-invariant). In complete analogy
with the Lie algebra case we would therefore like to have a formula
which relates the integral over $\bG$ to an integral over $\bT$.
In order to do that we will need a slightly more detailed understanding
of the conjugation action.

While it is true that any two maximal
tori are conjugate to each other (and hence all that follows is essentially
independent of the choice of maximal torus), it is not necessarily
true that the
centralizer $C(g)$ of an element $g \in \bG$ (i.e.~the
set of elements of $\bG$ commuting with $g$) is some maximal torus.
For example, for $g$ an element of the center $Z(\bG)$ of $\bG$ one
obviously has $C(g)=\bG$. However, the set of elements of $\bG$
whose centralizer is conjugate to $\bT$
is open and dense in $\bG$ and is called the set
$\bG_{r}$ of regular elements of $\bG$.

It follows that the conjugation map
\bea
\bG/\bT \times \bT_{r} &\ra& \bG_{r} \nonumber\\
(g,t) &\mapsto& \g t g \label{w1}
\eea
is a $|W|$-fold covering onto $\bG_{r}$. It is clear that the restriction
to $\bG_{r}$ is required here, because e.g.~an element $h$ of the center
$Z(\bG)$ of $\bG$ can be written as $\g h g$ for any $g\in \bG$ and thus
clearly constitutes a singular point for the above conjugation map.

\nul{The Weyl Integral Formula}

On $\bG$ and $\bT$ there
exist natural invariant Haar measures $dg$ and $dt$
normalized to $\int_{\bG}dg= \int_{\bT}dt = 1$. We use these to define
the spaces $L^{2}(\bG)$, $L^{2}(\bT)$, $L^{2}(\bG)^{G}$ (the subspace
of $L^{2}(\bG)$ consisting of conjugation invariant functions) etc.
For the purpose of
integration over $\bG$ we may restrict ourselves to $\bG_{r}$
(provided that the function $f$ is reasonably nice and not somehow
concentrated on the singular points) and we can
thus use (\ref{w1}) to pull back the measure $dg$ to $\bG/\bT \times \bT$.
To calculate the Jacobian, we need to know the infintesimal
conjugation action of $\bT$ (and only of $\bT$, see the
discussion of this issue in the Lie algebra case in section 2)
on $\bG/\bT$. Recall that corresponding to a choice of $\bT$
we have a direct sum decomposition of the Lie algebra
$\lg$ of $\bG$, $\lg = \lt \oplus \lk$, orthogonal with respect to the
Killing-Cartan metric on $\lg$. $\bG$ acts on $\lg$ via the adjoint
representation $\rm Ad$. This induces an action of $\bT$ which acts trivially
on $\lt$ and leaves $\lk$ invariant (the isotropy representation
${\rm Ad}_{\lk}$ of $\bT$ on $\lk$, the tangent space to $\bG/\bT$).
Therefore the Jacobian matrix is
\be
\dw(t) =  ({\bf 1}_{\lk}- {\rm Ad}_{\lk}(t)) \label{w3}
\ee
and one finds the Weyl integral formula
\be
\int_{\bG}\!dg\;f(g) =\trac{1}{|W|}\int_{\bT}\!dt\; \det\dw(t)
                    \int_{\bG/\bT}\!dg\;f(\g t g)\;\;.\label{w2}
\ee
In particular, if $f$ is conjugation invariant, this reduces to
\be
\int_{G}\!dg\;f(g) = \trac{1}{|W|}
           \int_{\bT}\!dt\; \det\dw(t) f(t) \;\;, \label{w4}
\ee
which is the version of the \wif\ which we will make use of later on.
Notice that the infinitesimal version of (\ref{w3}) is precisely the
determinant $\det{\rm ad}_{\lk}$ we encountered in the Lie algebra case, as
it should be.

As the complexified Lie algebra
$\lg_{\CC}$ splits
into $\lt_{\CC}$ and the one-dimensional eigenspaces $\lg_{\a}$ of the
isotropy representation, labelled by the roots $\a$ (see Appendix A), it
follows that the Jacobian (Weyl determinant) can be written as
\bea
\det({\bf 1}-{\rm Ad}_{\lk}(t)) &=&\prod_{\a} M_{\a}(t)\;\;,\nonumber\\
M_{\a}(t) &=& (1-e^{\a}(t))\;\;.\label{w7}
\eea
By decomposing the set of roots into positive ($\a > 0$) and negative roots,
this expression can also be written more explicitly as a product of sines
(see below for the formulae in the case of $SU(2)$ and $SU(3)$).

Another useful way of writing this
determinant is in terms of the so-called Weyl denominator.
Introducing the Weyl vector $\rho = \frac{1}{2}\sum_{\a > 0} \a$ and the
denominator $Q(t)$ of the Weyl character formula,
\be
Q(t)= \sum_{w\in W}\det(w)e^{w(\rho)}(t)\;\;,\label{w8}
\ee
a Weyl-odd (or anti-invariant) function on $\bT$,
one can write $\det\dw(t)$ as
\be
\det({\bf 1}-{\rm Ad}_{\lk}(t)) = Q(t)\overline{Q(t)} \;\;.\label{w9}
\ee
In (\ref{w8}), $\det(w)$ denotes the determinant of $w\in W$ regarded
as an orthogonal transformation on $\lt$ (alternatively, $\det(w)=-1$
if $w$ can be written as a product of an odd number of elementary reflections
along the walls of the Weyl chamber, and $\det(w)=1 $ otherwise).

As the above may have been somewhat technical and dry,
we will now illustrate various aspects and facets of the above in
some simple examples. We spell out the Weyl integral formula
explicitly for $SU(2)$ and $SU(3)$,
we indicate (in analogy with the procedure adopted
in the Lie algebra case) how it can be derived from the BRST and
Faddeev-Popov points of view, and we will say a little bit about characters.

\nul{$SU(2)$ and $SU(3)$}

We parameterize elements of $SU(2)$ and $\bT=U(1)$ as
\be
g = \mat{g_{11}}{g_{12}}{g_{21}}{g_{22}}\;\;,\;\;\;\;\;\;
t = \mat{e^{i\vf}}{0}{0}{e^{-i\vf}}\;\;.                     \label{w10}
\ee
The Weyl group $W=\ZZ_{2}\equiv S_{2}$ acts on $\bT$ as $t\mapsto t^{-1}$.
We use the trace to identify the Lie algebra $\lt$ of $\bT$ with its dual
and introduce the positive root $\a$ and the fundamental weight $\la$,
\be
\a = \mat{1}{0}{0}{-1}\;\;,\;\;\;\;\;\;\la =\trac{1}{2}\mat{1}{0}{0}{-1}\;\;,
\ee
satisfying the relations
\be
\tr \a^{2} = 2 \;\;,\;\;\;\;\;\;\tr \a\la = 1\;\;,\;\;\;\;\;\;
\rho = \trac{1}{2}\a = \la\;\;.
\ee
Later on we will find it convenient to parameterize elements of $\bT$ in terms
of weights. Thus, we write
$t= \exp i\la\phi$, where $\f$ is related to $\vf$ by $\f=2\vf$.
Then the expression $\exp(\a)(t)$
entering (\ref{w7}) becomes $\exp(\a)(t) = \exp i\f$, and
the Weyl denominator (\ref{w8}) and the determinant (\ref{w3}) are
\bea
Q(t)&=& 2i\sin\f/2\;\;,\nonumber\\
\det\dw(t) &=& 4 \sin^{2}\f/2\;\;.                            \label{w11}
\eea
Hence the \wif\ for class functions is  (with $f(\f)\equiv f(\exp i \la\f)$)
\bea
\int_{\bG}\!dg\;f(g) &=& \frac{1}{2}\int_{0}^{4\pi}\!\frac{d\f}{4\pi}4
\sin^{2}(\f/2) f(\f)\nonumber\\
&=& \frac{1}{2\pi}\int_{0}^{4\pi}\!d\f\; \sin^{2}(\f/2) f(\f)\nonumber\\
&=& \frac{1}{\pi}\int_{0}^{2\pi}\!d\f\; \sin^{2}(\f/2) f(\f)\;\;. \label{w12}
\eea
Here the last line follows e.g.~from writing $2\sin^{2}(\f/2)=
1-\cos \f$ and
is a useful reformulation because it effectively incorporates the action
of the Weyl group.

For $G=SU(3)$, there are three positive roots which we
take to be
\be
\a_{1}={\rm diag}(1,-1,0)\;\;,\;\;\;\;\a_{2}={\rm diag}(0,1,-1)\;\;,
\;\;\;\;\a_{3}=\a_{1}+\a_{2}\;\;,
\ee
with $\a_{1}$ and $\a_{2}$ simple. These determine the corresponding
fundamental weights $\la^{k},\; k=1,2$ with
\be
\tr \a_{k}\la^{l} = \d_{k}^{\,l}
\ee
to be
\be
\la^{1} = {\rm diag}(2/3,-1/3,-1/3)\;\;,\;\;\;\;
\la^{2}={\rm diag}(1/3,1/3,-2/3)\;\;.\label{su3}
\ee
Writing, as above, $t\in \bT$ as $t=\exp i \f$ with $\f = \la^{k}\f_{k}$,
the Weyl determinant becomes (modulo a factor of $4^{3}$)
\bea
\det\dw(t) &\sim& \prod_{\a>0} \sin^{2}(\a(\f)/2) \nonumber\\
   &=& \sin^{2}(\f_{1}/2) \sin^{2}(\f_{2}/2) \sin^{2}(\f_{1}+\f_{2}/2)\;\;.
\eea

\nul{Faddeev-Popov Derivation}

As in the Lie algebra case
these formulae can be obtained \`a la Faddeev-Popov by `gauge fixing' the
non-torus part of $g$ to zero (i.e.~by imposing $g\in\bT$ as a gauge
condition). This amounts to inserting $1$ in the form
\be
1=\trac{1}{|W|}\int_{\bG/\bT}\!dh\;\int_{\bT}\!dt\;\d(h^{-1}gh t^{-1})
\det\dw(t)
\label{w5}
\ee
into the integral on the lhs of (\ref{w2}) and performing the integral
over $g$,
\bea
\int_{\bG}\!dg\;f(g) &=&
\trac{1}{|W|}\int_{\bG}\!dg\;\int_{\bG/\bT}\!dh\;\int_{\bT}\!dt\;
\d(h^{-1}gh t^{-1})\det\dw(t)\nonumber\\
&=&\trac{1}{|W|}\int_{\bT}\!dt\;\det\dw(t)\int_{\bG/\bT}\!dh\;f(h^{-1}th)
\;\;.
\eea
The Faddeev-Popov determinant $\det\dw(t)/|W|$ can then be
obtained either by calculation of the corresponding Jacobian (as we did
above) or more directly from the (BRST) variation of the condition $g\in\bT$.
In the $SU(2)$ case this amounts to fixing the gauge
$g_{12}=g_{21}=0$. Since
infinitesimally $g_{12}$ transforms under conjugation as ($a\in\lg$)
\be
\d g = [g,a] \Rightarrow \d g_{12} = 2i a_{12} \sin\vf\label{w13}\;\;,
\ee
the resulting Faddeev-Popov determinant is just (\ref{w11}), while the
additional factor of $1/2$ accounts for the residual gauge freedom
(conjugations leaving $\bT$ invariant).

Quite generally, it is easy to see that (with the notation introduced above)
the ghost contribution to the `action' is
\be
\sum_{\a>0}\left[\bar{c}^{\a}*M_{-\a}c^{-\a}
    + \bar{c}^{-\a}*M_{\a}c^{\a}\right]     \;\;,
\ee
leading to the determinant $\prod_{\a}M_{\a}$.

\nul{Characters}

It is a consequence of the Peter-Weyl theorem that
the space of class functions on a compact Lie group $\bG$ is spanned
by its irreducible characters, i.e.~by the traces in the unitary irreducible
representations $R\in\hat{\bG}$ of $\bG$,
\bea
L^{2}(\bG)^{G} &=& {\rm span}_{\CC}\{\c_{R}\;,\;R \in \hat{\bG}\}
\;\;,\nonumber\\
\c_{R}(g) &=& \tr\left(R(g)\right)\;\;.
\eea
These characters are orthonormal with respect to the $L^{2}$ inner product
on $\bG$,
\be
\int_{\bG}\!dg\,\c_{R}(g)\overline{\c_{S}(g)}=\d_{R,S}\;\;,
\ee
(where, because of unitarity, $\overline{\c_{S}(g)} = \c_{S}(\g)$)
and hence any class function can be expanded in a `Fourier' series in
the functions $\c_{R}$. As class functions, the characters themselves
are also uniquely determined by their restriction to $\bT$ (where they
are Weyl invariant) and the Weyl character formula expresses these
as ratios of two $W$-odd functions on $\bT$. If $\mu$ is the highest
weight of the representation $R$, then $\c_{\mu}\equiv \c_{R}$ is given
by
\be
\c_{\mu}(\exp i\f) = \frac{A_{\mu+\rho}(\exp i\f)}{A_{\rho}(\exp i\f)}\;\;,
\ee
where the functions
\bea
A_{\nu}(\exp i\f) &=& \sum_{w\in W}\det(w)\ex{i(w(\nu),\f)}\;\;,\nonumber\\
A_{\rho}(\exp i\f) &\equiv& Q(\exp i\f)\;\;,
\eea
form a basis for the space of $W$-odd functions on $\bT$. For $SU(2)$,
where the Weyl group consists of only two elements, these
are just the sine functions on the circle.
For example, for the spin $j$ representation of $SU(2)$ (with highest
weight $\mu_{j}=2j\la$, $\la$ the fundamental weight of $SU(2)$), one finds
\be
\c_{\mu_{j}}(\exp i\la\f) = \frac{\sin\trac{(2j+1)\f}{2}}{\sin\trac{\f}{2}}
\;\;.
\label{char}
\ee
The Weyl dimension
formula we used in section 2.3 can be extracted from the Weyl character
formula by (carefully) evaluating $\c_{\mu}$ on the unit element $\bf 1$,
as $\c_{\mu}({\bf 1}) = d(\mu)$. E.g.~in the above example one finds
\be
\lim_{\f\ra 0} \c_{\mu_{j}}(\exp i\la\f) = 2j+1 = d(\mu_{j})\;\;.
\ee
The Weyl integral formula translates the orthonormality of the
characters on $\bG$ into the orthonormality of the functions $A_{\la+\rho}$
on $\bT$ with respect to the Haar measure $dt/|W|$,
\bea
\d_{\la,\mu}&=&
\int_{\bG}\!dg\,\c_{\la}(g)\c_{\mu}(\g)\nonumber\\
&=& \trac{1}{|W|}\int_{\bT}\!dt\,\det\dw(t) \c_{\la}(t)\c_{\mu}(t)\nonumber\\
&=&  \trac{1}{|W|}\int_{\bT}\!dt\,Q(t)\bar{Q}(t)\frac{A_{\la+\rho}(t)}{Q(t)}
      \frac{A_{\mu + \rho}(t^{-1})}{\bar{Q}(t)}\nonumber\\
&=&  \trac{1}{|W|}\int_{\bT}\!dt\,A_{\la+\rho}(t) \bar{A}_{\mu+\rho}(t)\;\;.
\eea

\subsection{Abelianization of the ${\bf G/G}$ Model}

Our aim in this section will be to calculate explicitly the partition function
of the $G/G$ model on a closed Riemann surface as well as the correlation
functions of the traces $\tr g$.
Applied to the path integral of the $G/G$ model, the \wif\
permits one to effectively reduce the path integral to that of an Abelian
theory which can be exactly calculated. The treatment in this section
will follow closely that given in \cite{btver}.

As a consequence of the gauge symmetry (\ref{wz3}) of the $G/G$ action,
the functional of $g$ that
one obtains after having performed the path integral over the gauge
fields is locally and pointwise conjugation invariant,
\be
{\cal F}(g)\equiv\int DA \exp(ik S_{G/G}(g,A)) = {\cal F}(h^{-1}gh)
\label{g1}\;\;.
\ee
This shows that there is enough gauge freedom in the theory to conjugate
$g$ into the maximal torus $\bT$ (i.e. to impose the gauge condition
$g\in \bT$). As in the case of Yang-Mills theory, there may
be obstructions to doing this globally with continuous gauge
transformations, and this will once again give rise to a summation
over non-trivial torus bundles as will be explained in \cite{btwif}.

Hence we can (formally) use the \wif\ (\ref{w2}) in its
strong form (\ref{w4}) pointwise to reduce the
path integral over the group valued fields to one over fields
taking values in $\bT$. The Faddev-Popov determinant arising from this gauge
choice will be a functional version of
the Weyl determinant $\det\dw(t)$ we encountered in (\ref{w3}) above. Hence
\be
\int Dg {\cal F}(g)=
\int Dt DA\,{\rm Det}\,({\bf 1}-{\rm Ad}_{\lk}(t))
\exp(ik S_{G/G}(t,A))\;\;,\label{g2}
\ee
where $\Det$ denotes a functional determinant.
We will define and evaluate
these and other determinants arising in the following in Appendix B.

The replacement of $g$ by $t$ leads to a
significant simplification of the rather uninviting action
(\ref{diff}) of the $G/G$ model. In particular, we will see that the
$\lt$ and $\lk$ components of the gauge field $A=A^{\lt}+A^{\lk}$
decouple from each other and that the latter can easily be integrated out
to leave one with an effective Abelian theory. To obtain an explicit form
for the action we expand (as in the case of Yang-Mills theory) the
gauge field $A^{\lt}$  and the torus valued field $t$ in terms of simple
roots $\{\a_{l},\;l=1,\dots, n-1\}$ and their dual fundamental weights
$\{\la^{l}\}$ respectively,
\bea
A^{\lt}&=&i\a_{l}A^{l}\nonumber\\
t&=&\exp i\f\;\;,\;\;\;\;\;\;\f = \f_{l}\la^{l}\;\;.\label{para}
\eea
It is clear from this description that the $\f_{l}$ are compact
scalar fields. E.g.~in the case of $SU(3)$ it follows from the
explicit form (\ref{su3}) of the weights that
\be
0\leq2\f_{1}+\f_{2}\leq 6\pi\;\;,\;\;\;\;\;\;0\leq\f_{2}-\f_{1}\leq 6\pi
\;\;.
\ee
In terms of fields $X_{1}$ and $X_{2}$ related to $\f_{1}$ and $\f_{2}$
by $\f_{1} = X_{1}-X_{2}$ and $\f_{2} = X_{1}+ 2 X_{2}$ this is just the
`standard' torus $0\leq X_{l}\leq 2\pi$.
However, as we will eventually still want to mod out by the Weyl group,
the range of $\f$ will be restricted further to $\bT/W$ or, rather, to a
fundamental domain for the action of $W$ on $\bT$, and we will return to
this issue below.

We will now take a look in turn at the various contributions to the
action $S_{G/G}(t,A)$: the kinetic term $S_{0}(t)$ (\ref{3}), the
WZ term $\Gamma(t)$ (\ref{4}), and the terms in $S_{/G}(t,A)$
(\ref{6},\ref{diff}) linear and quadratic in $A$ respectively.

The kinetic term $S_{0}(t)$ obviously reduces to the (more or less)
standard kinetic term
\be
S_{0}(t) =
\trac{1}{4\pi}\int_{\S}  \la^{kl}
            \dz\f_{k}\dzb\f_{l}\label{g3}
\ee
for compact bosons. Here $\la^{kl}={\rm tr}(\la^{k}\la^{l})$ is
the inverse of the Cartan matrix. As this matrix is not diagonal,
there will be off-diagonal couplings among the scalar fields $\f_{k}$
(hence the `more or less' above).

As a `topological' term, the WZ term $\Gamma(t)$ turns out to depend
only on the
winding numbers of the field $\f$ along the cycles of $\S$. While this
is perhaps not in itself unusual, the surprise
is that a contribution from the WZ term arises at all, since
the WZ term for an Abelian group is identically zero. The reason
for the appearance of this contribution
is, that maps from $\S$ to $\bT$ with non-trivial windings cannot
necessarily be extended to the interior $N$ of $\S$ {\em within} $\bT$,
as some (half) of the non-contractable cycles of $\S$ become contractable
in the `handlebody' $N$ with $\del N = \S$.

To have a concrete example
in mind, take $\S$ to be a torus and let $N$ be the solid torus in which
one of the two originally non-contractable cycles of $\S$, say the $a$-cycle,
has become contractable. A map from $\S$ to $\bT$ which can be extended
to a map from $N$ to $\bT$ then necessarily has zero winding number along the
$a$-cycle. Conversely, a map with non-trivial winding number along that cycle
cannot be extended to a map from $N$ to $\bT$. It is of course
perfectly possible to start off with a map $g$ from $\S$ to $\bG$ (with an
extension $g_{N}$ to $N$ - recall the notation introduced after (\ref{4}))
which, when
conjugated into a map $t_{N}$ such that on the boundary it takes values in
$\bT$ (which we know is possible), has a non-trivial winding around that cycle.
What the discussion above shows, however, is that the range of $t_{N}$ in the
interior of $N$ cannot lie entirely in $\bT$. It is this that
allows the WZ term $\Gamma(t_{N})\equiv \Gamma(t)$ to be non-zero.

The general form of this term is \cite{gacs}
\be
\Gamma(t) =
\int_{\S} \mu^{kl}\,d\f_{k}\,d\f_{l} \;\;,
\ee
where $\mu^{kl}$ is some antisymmetric matrix. As this is almost exact
(were it not for the compactness of the $\f_{k}$), it is clear from this
expression that it is invariant under variations of the fields and can hence
depend only on their global properties, the winding numbers.
As we will show below
(cf.~the discussion after (\ref{bf2act}))
that the non-trivial winding sectors do not contribute to the partition
function, we will not have to be more precise about this term here.

We now come to the part of the action involving the gauge fields.
In $S_{/G}(t,A)$, the contributions from the $\lt$ and $\lk$
components $A^{\lt}$ and $A^{\lk}$ of the gauge field $A$ are neatly
separated so that it is easy to perform the path integral over the $A^{\lk}$,
leaving behind an effective Abelian theory. In fact,
because $\lt$ and $\lk$ are orthogonal to each other with respect to the
invariant scalar product (trace), $A^{\lt}$
will obviously not appear in the term
$\az\azb - t^{-1}\az t\azb$ which becomes simply
\be
\tr \az\azb - t^{-1}\az t\azb =
\tr \az^{\lk}({\bf 1}-{\rm Ad}_{\lk}(t))\azb^{\lk}
\;\;.\label{g4}
\ee
Thus the $A^{\lk}$ integral will give rise to a determinant that {\em
formally} cancels against the Faddeev-Popov determinant in (\ref{g2}).
Of course, as in Yang-Mills theory,  this is not quite correct,
as the Faddeev-Popov determinant is a scalar determinant while in (\ref{g4})
we have an operator acting on one-forms. Hence certainly the zero modes will
leave behind a finite dimensional determinant. Furthermore, the determinants
should be properly regularized, and we will evaluate this ratio of
determinants in
Appendix B. Suffice it to say here that this gives rise to
the shift $k\ra k+h$. In fact, the residual finite dimensional determinant
$\det^{1-g}\dw(t)$ and the shift will arise simultaneously as the
gravitational and gauge field contributions to the chiral anomaly.

On the other hand, only the $A^{\lt}$ will contribute to
the terms of the form $\azb t^{-1}\dz t$ and $\az \dzb t \,t^{-1}$
(cf.~equation (\ref{5})),
\be
-\trac{1}{2\pi}\int_{\S} (\az\dzb t \,t^{-1} - \azb t^{-1}\dz t) =
\trac{1}{2\pi}\int_{\S}( \az^{l}\dzb\f_{l} - \azb^{l}\dz\f_{l})
\;\;.\label{g5}
\ee
Thus, putting everything together we see that upon Abelianization
(going to the gauge $g\in T$) and elimination of the $\lk$ components
of the gauge field we are left with a theory described by
the simple Abelian action
\be
\int_{\S}\left(\trac{1}{4\pi}\la^{kl} \dz\f_{k}\dzb\f_{l}
+\trac{1}{2\pi}(\az^{l}\dzb\f_{l} - \azb^{l}\dz\f_{l})
+\mu^{kl}\,d\f_{k}\,d\f_{l}\right) \;\;,
\label{abs}
\ee
and the non-linear $\f$ measure $\det^{1-g}\dw(t)$.

\subsection{Relation with BF Theory}

While (\ref{abs}) is as far as Abelianization will get us,
this action can still be put into a manifestly topological form which
reveals its relation to the topological BF theory discussed in the
first part of these lectures.

First, we observe that we can eliminate the kinetic term (\ref{g3})
for the scalar fields altogether from (\ref{abs}) by
a shift of the gauge field,
\be
A^{\lt}\ra A^{\lt} + \trac{1}{4}*d\f^{\lt}\;\;, \label{g6}
\ee
(note that this is not a gauge transformation)
leaving us with the even simpler action
\be
\trac{1}{2\pi}\int_{\S}\tr A\,d\f
+\int_{\S}\mu^{kl}\,d\f_{k}\,d\f_{l} \;\;.
\label{act2}
\ee
As we will see presently that only the constant modes of $\f$ contribute
to the path integral we could just as well carry the term
(\ref{g3}) along until the end. And while this would avoid the seeming
nuisance of a metric dependent field redefinition, it is nicer to
work with the action (\ref{act2}) because of its resemblance to
other topological gauge theories in two dimensions.

We would now like to integrate by parts in the first term of
(\ref{act2}) to put it into the form of the action of a BF theory,
whose action in $2d$, we recall, is
\be
S_{BF}=\trac{1}{2\pi}\int_{\S}\tr BF_{A}\;\;,\label{bf2act}
\ee
where $B$ is an ordinary (non-compact) scalar field.

At first, the
compactness of $\f$ may cast some doubt on this procedure since,
with $\f$ being an `angular variable', $d\f$ is not necessarily exact. One
would therefore expect to pick up `boundary' terms from the monodromy of $\f$.
The following argument shows that in the $G/G$ model the non-trivial
winding sectors of
these fields do not contribute to the partition function: as it is only
the harmonic modes of $A$ that couple to the non-exact (winding) parts
of $d\f$ and $A$ appears nowhere else in the action, these harmonic modes
act as Lagrange multipliers setting the non-zero
winding modes of $\f$ to zero. Thus, upon integration over the harmonic modes
of $A$ we can
indeed integrate by parts in (\ref{act2}) with impunity, the residual WZ term
also disappears from the action,  and (with the understanding that
the harmonic modes of $A$ no longer appear) we thus arrive at the
BF like action
\be
S_{\f F} = \trac{1}{2\pi} \int_{\S}\f_{l}F^{l}\;\;.\label{g7}
\ee
Here $F^{l}=dA^{l}$ is the curvature of the Abelian gauge field $A^{l}$.

One of the important differences between this theory and
the ordinary BF models is of course, that here
the scalar fields $\f_{l}$ are compact which implies that the
integral over them will not simply produce a delta function onto flat
connections as is the case in the non-compact BF theories. We will make
some more comments on the relation between these two theories below.

Anticipating the results of Appendix B, we have thus deduced that
the $G/G$ model on
$\S$ (and hence Chern-Simons theory on $\SS$) is equivalent to
an Abelian topological field theory,
\be
Z_{\S}(S_{G/G}) = \int D[\phi,A] \exp(i(k+h)\,S_{\phi F}(\phi,A))\;\;,
\label{phif1}
\ee
with action  and measure given by
\bea
S_{\phi F}(\phi,A)&=&\trac{1}{2\pi}\int_{\S}\tr \phi F_{A}\;\;,
\label{phif2}\\
D[\phi,A] &=& D\phi\,DA\,\det({\bf 1}-{\rm Ad}(e^{i\phi}))^{1-g}
\label{phif3}
\eea
respectively.

The action (\ref{phif2}) is obviously a `compact' counterpart of
BF theory, defined in any dimension $n$ by (\ref{bf2act}) with $B$ a
Lie algebra valued $(n-2)$-form. In fact, by comparing with the
Abelianization of non-Abelian BF theory, one again recognizes the latter
to be a (tangent-space) linearization of the $G/G$ model.
It will be
useful to keep in mind the following differences between the compact
and non-compact models in two dimensions:

\begin{enumerate}

\item As we have seen in section 2,
in two dimensions (and, with some {\em caveat}
also in general, see \cite{btbf1}) the integral over $B$ simply imposes
the delta function constraint  $F_{A}=0$, so that the partition function
calculates the volume of the moduli space of flat connections, with
measure given by the Ray-Singer torsion. In $n=2$ this measure coincides
with the symplectic measure \cite{ewym} and hence, with proper
normalization, the partition function is the symplectic volume of
$\MF$. The compactness of $\f$ in (\ref{phif2}) on the other hand
implies that the partition function is no longer a simple delta
function but some deformation thereof. In fact, in terms of a
suitably chosen mode expansion (spectral representation of the
delta function) one finds that sufficiently high modes of
the delta function are cut off due to the compactness of $\f$.

\item Note also that, in the case of BF theories, any prefactor (coupling
constant) like $k$ in the path integral can be absorbed by a rescaling of
$B$, so that the result is essentially independent of $k$.
This is something that, due to the compactness of $\f$, cannot be
done in the action $S_{\f F}$, as a rescaling of $\f$ would change
its radius. We thus expect the partition function (and hence that of \CS\
theory) to depend in a much more subtle manner on $k$, something that
is indeed borne out by the result, the Verlinde formula. One would,
however, expect the large $k$ limit of this result to agree with the
partiton function of BF theory since, by rescaling, the large $k$
limit corresponds to a larger and larger radius of $\f$. This can indeed
be verified and is in
agreement with the expectation that in the semi-classical limit
of \CS\ theory the dimension of the Hilbert space is equal to the
volume (number of cells) of phase space. To see this directly,
one can argue (see \cite{ewym}) that in the stationary phase
approximation (large $k$ limit) of \CS\ theory on $\SS$ the dominant
contributions come from flat connections on $\S$ (with multiplicity the
order of the center of $\bG$, coming from the possible holonomies around the
$S^{1}$), while the Chern-Simons action for connections of the form
$\unA = Bdt + A$, with $B$ and $A$ a scalar and a connection on $\S$,
reduces to the BF action on $\S$.
We will confirm this reasoning below by deriving
the BF partition function for surfaces with marked points from the large $k$
limit of the $G/G$ partition function (equivalently, the Verlinde
formula).


\item Finally we note that, as
a compact scalar field may be regarded as a non-compact scalar
field modulo a (one-dimensional) lattice, one
way to relate the compact and non-compact models is the
following. The non-compact Abelian BF action (or rather $\exp ik S_{BF}$)
enjoys the invariance $B\ra B + \gamma, \gamma\in I$, where $I$ is a
suitable lattice (in fact, the integral lattice of $SU(n)$), since
\be
\int_{\S}F_{A} \in 2\pi\ZZ\;\;. \label{chern2}
\ee
As this is just some global symmetry of the action,
all one can expect is to find
it unitarily represented on the Hilbert space of the theory.
However, were one to promote this symmetry of the action to a
`large' gauge invariance
(by adding the instruction to mod out by this symmetry) one would indeed be
dealing with a compact scalar field, the theory now being described by
the invariant subsector of the non-compact model. Of course, the measure
of the $\phi F$ theory (the carrier of the information on the non-Abelian
and non-linear origin of the $\phi F$ theory (\ref{phif1}-\ref{phif3}))
remains different from the linear measure of Abelian BF theory and
the Lie algebra Weyl determinant measure one obtains upon Abelianization
of the non-Abelian BF model.

\end{enumerate}

\subsection{Evaluation of the Abelian theory - The Verlinde Formula}

In this section we will evaluate the partition function
\be
Z_{\S}(S_{\phi F},k) = \int\! D\f\,DA\,
\det({\bf 1} - {\rm Ad}(e^{i\f}))^{\c(\S)/2}
\exp\left(\trac{i(k+h)}{2\pi}
\int_{\S}\tr\f F_{A}\right)\;\;,\label{zphif}
\ee
which, as we have seen above, is equal to
the partition function of the $G/G$ model on $\S$. To a large extent this
can be done exactly as in (solution 1 of) the evaluation of the Yang-Mills
partition function in section 2, the only differences being that here
$\f$ is compact and there is no term quadratic in $\f$.\footnote{An alternative
derivation, based on the use of the trivializing map (change of variables)
$A\ra F_{A}$, is given in \cite{btver}.}
Rewriting, as in
(\ref{pdf}), the periodic delta function as an infinite sum over the weight
lattice, one obtains
\bea
Z_{\S}(S_{\phi F},k) &=&
\sum_{\lambda\in\Lambda}\int\! D\f\,
\det(\dw(e^{i\f}))^{\c(\S)/2}
\delta(\trac{k+h}{2\pi}\f - \lambda) \nonumber\\
&=&
\prod_{k=1}^{r}\sum_{n_{k}}\int\! D\f^{k}\,
\det(\dw(e^{i\f}))^{\c(\S)/2}
\delta(\trac{k+h}{2\pi}\f_{k} - n_{k})
\;\;.\label{zphif4}
\eea
The first thing to note is that this equation implies in particular
that only the constant modes of $\f$ contribute to the partition
function. This has two important consequences, namely a) that
the dilaton term (see appendix B) turns into the metric independent
$(g-1)$'th power of the Weyl determinant, and b) that had we
carried around the
kinetic term (\ref{g3}) for the compact scalars until now (instead of
eliminating it by the shift (\ref{g6})), it would disappear now.

We see that we have
been led to a sum over the weight lattice, whose
summation range will be restricted by the compactness of $\f$.
This will turn the sum over all representations (which we obtained
in the case of Yang-Mills theory) into a sum over
the highest
weights of integrable representations at level $k$. To make this
explicit, it is
convenient to restrict the integration range for $\f$ to a
fundamental domain of the action of the Weyl group $W$ on $\bT$
(which is the only piece of gauge freedom we have not yet fixed).
As $W$ is a finite group and the integral (\ref{zphif4}) is manifestly
$W$-invariant, the result will be the same as dividing the
integral by $|W|$, but not necessarily manifestly so, as the sum
will then extend over all the weights in the $W$-orbits of
highest weights of integrable representations.
We shall not worry about the overall normalization of the
path integral (see \cite{btver}).

\nul{$SU(2)$ Partition Function}

Let us first see what happens in the case of $SU(2)$ before dealing with
$SU(n)$ in general.
In this case, $\f$ is a single compact scalar, $r=1$ and
$\Lambda\sim{\bf Z}$ in (\ref{zphif4})
and the Weyl determinant is $4\sin^{2}(\f/2)$. By the
action of the Weyl group the range of $\f$ is cut down from $[0,4\pi)$
to $[0,2\pi]$ and it is convenient to use the form of the Weyl integral
formula given in the last line of (\ref{w12}). Thus, for $SU(2)$ we
have
\be
Z_{\S}(S_{\f F},k) = \sum_{n=-\infty}^{+\infty}\int_{0}^{2\pi}\!d\f\,
\sin^{2-2g}(\f/2)\,\delta(\trac{k+2}{2\pi}\f - n)\;\;. \label{zphif5}
\ee
In particular, only certain discrete values of $\f$ contribute to the
path integral and due to the compactness of $\f$ only a finite number
of $n$'s give a non-vanishing contribution. Ignoring the boundary values
$n=0$ and $n=k+2$ for a moment (we will come back to them below) we
see that the allowed values of $\f$ are
\be
\f = \frac{2n\pi}{k+2}\;\;,\;\;\;\;\;\; n = 1,\ldots, k+1\;\;.\label{alp}
\ee
These points are in one-to-one correspondence with the $k+1$ integrable
representations of the $SU(2)$ WZW model at level $k$ and we see
that, up to normalization, the partition function is
\be
Z_{\S}(S_{\f F},k) = \sum_{n=1}^{k+1}\sin^{2-2g}(\trac{n\pi}{k+2})\;\;.
\label{zphif6}
\ee
This compares favourably with the Verlinde formula
\be
\dim V_{g,k} = (\trac{k+2}{2})^{g-1}\sum_{j=0}^{k}
\left(\sin^{2}\trac{(j+1)\pi}{k+2}\right)^{1-g}\;\;.\label{ver3}
\ee
for the dimension of the space of conformal blocks of the level $k$
$SU(2)$ WZW model on a genus $g$ surface.

\nul{Correlation Functions and Marked Points}

The calculation of a correlator of $s$ traces of $g$ is equally
straightforward.
Let $\c_{l}(h)$ be the character (trace) of $h\in SU(2)$
in the $(l+1)$-dimensional representation of $SU(2)$. We will only
consider integrable representations, $l\leq k$.
As the characters are conjugation invariant, we need to know them only on the
maximal torus, where they can be expressed as (see \ref{char})
\be
\c_{l}(\f)\equiv\c_{l}(e^{i\f}) = \frac{\sin(l+1)\f/2}{\sin\f/2}\;\;.
\label{f2}
\ee
Then the above reasoning leads to
(\ref{zphif4}) with an insertion of $s$ characters in the
form (\ref{f2}). And, just as in the case without insertions, evaluation
of the delta function will lead to a sum over the discrete
allowed values of $\f$ and one finds
\be
\langle\prod_{i=1}^{s}\c_{l_{i}}(\f)\rangle_{g} =
\sum_{j=0}^{k}\left(\sin\trac{(j+1)\pi}{k+2}\right)^{2-2g-s}\prod_{i=1}^{s}
\sin\trac{(j+1)(l_{i}+1)\pi}{k+2}\;\;,\label{f3}
\ee
which, up to normalization, agrees with the Verlinde formula (\ref{ver2})
for the dimension of the vector space $V_{g,s,k}$.

\nul{The $k\ra\infty$ Limit}

In order to confirm the arguments concerning the equivalence of the
large $k$ limit of the $G/G$ model with BF theory,
let us now take a look at the large $k$ limit of the results
we have obtained for the partition function and the correlators. As $k
\ra \infty$ we wish to extract, from (\ref{ver3}), the part that grows like
$k^{3g-3}$. There are two terms in the sum that contribute.  For fixed
$j$ and large $k$ ($j<<k$) one obtains
\be
\sin{}^{2-2g}\left(\frac{(j+1)\pi}{k+2}\right) \sim
\left(\frac{k}{(j+1)\pi}\right)^{2g-2} \, .
\ee
The other region that contributes (equally) is fixed $k-j$ such that
$k-j <<k$. Putting the pieces together we find that as $k \ra \infty$
\be
\dim{V_{g,k}} \sim 2 \frac{k^{3g-3}}{2^{g-1} \pi^{2g-2}}
\sum_{n=1}^{\infty} \frac{1}{n^{2g-2}} \, ,
\ee
in complete agreement with the results of section 2.5. A similar
analysis of the large $k$ limit of (\ref{ver2}) reproduces the partition
function for the Riemann surface with marked points.

\nul{Discrete Characters}

Parenthetically we want to point out an interesting feature of the
above calculation, namely
that the characters only ever receive contributions form those
values of $\f$ where the classical characters satisfy the quantum fusion
rules. Less cryptically this means that, defining $\c_{l}^{(j)}$ by
\be
\c_{l}^{(j)} = \c_{l}(\f=\trac{2\pi(j+1)}{k+2}) \;\;,
\ee
one has
\be
\c_{l}^{(j)}\c_{m}^{(j)} = N_{lmn}\c_{n}^{(j)}\;\;,
\ee
where the fusion coefficients $N_{lmn}$ are given by the
three-point function on the sphere,
\be
\langle\c_{l}(\f)\c_{m}(\f)\c_{n}(\f)\rangle_{g=0} =
\sum_{j=0}^{k}\frac{S_{jl}S_{jm}S_{jn}}{S_{j0}}\equiv N_{lmn}\;\;.
\ee

\nul{The Non-Regular Points}

As in the case of Yang-Mills theory
we need to come to terms with the singular points of the gauge field
determinant. Here they are located at $\f=0$ and $\f=2\pi$,
coresponding to $n=0$ and $n=k+2$ in (\ref{alp}) and arising as
the boundary points of the reduced $\f$-range $[0,2\pi]$. The first
thing to note is, that these values correspond to the connections
on the circle with holonomy group $\{1\}$ (the trivial connection)
and $\{1,-1\}$ respectively. As such they are the most reducible
connections on the circle and require a special treatment in the
path integral. This can also be seen from our use of the Weyl integral
formula which, strictly speaking, only covers the regular elements
of $\bG$ or $\bT$, i.e.~excludes precisely the two special values
of $\f$ (for which the Weyl determinant vanishes).

The usual procedure would be
to either declare their contributions to be zero because of ghost zero modes
or to ignore these singular points.
Technically, this can be achieved either by adding a mass term for the
gauge fields (as we did in section 2) or
by choosing the integration range for
$\f$ to be $[\eps,2\pi - \eps]$ and taking the limit $\eps\ra 0$.
This also takes care of the problem that these boundary values give rise
to infinities in the partition function in genus $g>1$ (as may be
seen from (\ref{zphif6})) and any other method of regulating these
infinities would also amount to ignoring these contributions.
We can take the attitude that the WZW models are defined by integrating
over fields with values in $\bG_{r}$. Such configurations are dense in
the space of fields and the path integral is naturally regularized
by the restriction to $\bG_{r}$.

As this procedure may nevertheless seem somewhat ad hoc, we want to
point out that there is also another reason for `dropping' the boundary values
and (more generally) the points on the boundary of the Weyl alcove. Namely,
as is well known there is a quantization ambiguity in Chern-Simons
theory (see e.g. \cite{axcs,imcs}), corresponding to the option to work
with $W$-even or $W$-odd wave functions. While both of these appear to lead
to perfectly unitary quantizations of \CS\ theory (in genus one),
it is only the latter
which turns out to be related to the current blocks of $G_{k}$ WZW models.
In particular, for $SU(2)$ this forces the wave functions to vanish at
$\f =0$ and $\f = 2\pi$. And, while the differences between the two
alternatives are quite significant in general, for the
purposes of calculating the partition function they indeed only amount to
including or dropping the boundary values. Again we want to stress, however,
that we would like to see this prescription come out of the theory itself.

\nul{$SU(n)$ Partition Function}

We now turn to $SU(n)$.
The only complication that arises for $n>2$ is that we have
to prescribe a fundamental domain for the action of the Weyl group on
the torus $\bT = U(1)^{n-1}$. Alternatively, we are looking for a
fundamental domain of the action on $\lt$ of the semi-direct product of the
integral lattice (acting via translations) with the Weyl group (acting
via reflections). The advantage of this reformulation is that one now
recognizes this as a fundamental domain for the affine Weyl group (for
simply connected groups the integral lattice and the coroot lattice
coincide), which is known as a Weyl alcove or Stiefel chamber. In
particular, given such an alcove ${\bf P}$, we obtain a refinement of the
covering conjugation map (\ref{w1}) to a universal covering
\cite[Prop.~7.11]{btd}
\bea
\bG/\bT \times {\bf P} &\ra& \bG_{r}\;\;,\nonumber\\
(g,X)                  &\ra& \g \exp(X)\, g\;\;.
\eea
Hence this is an isomorphism if $\bG$ is simply connected and therefore
$\bf P$ is precisely the integration domain we require in the Weyl integral
formula instead of $\bT$ if we want to mod out by the Weyl group explicitly.

For $SU(n)$ such a Weyl alcove is determined by
$\a_{l}>0$ (fixing a Weyl chamber) and the one additional condition
$\sum\a_{l} < 2\pi$. As the fundamental weights are dual to the simple
roots, this amounts to the following conditions on
the integration range of $\f$:
\be
{\bf P} = \{\f_{l}:\,\f_{l}>0\,,\,\sum_{l=1}^{r=n-1}\f_{l}<2\pi\}\;\;.
\ee
For $SU(3)$ this can be checked directly, using e.g.~the action of the
Weyl group given in Appendix A, while in the general case it
is advisable to consult one's favourite textbook on group theory.

Introducing this constraint on $\f$
into the path integral (\ref{zphif4}), one
finds that only those weights ($r$-tuples of integers) contribute
to the partition function which satisfy $n_{l}>0$ and $\sum n_{l} < k+n$,
i.e.~the allowed values of $\f$ are
\be
\f_{l} = \frac{2\pi n_{l}}{k+n}\;\;,\;\;\;\;\;\;n_{l}>0\;\;,\;\;
\sum n_{l} < k+n\;\;.
\ee
Again these are in one-to-one correspondence with the integrable
representations of the $SU(n)$ WZW model at level $k$ and, up
to an overall normalization, the partition function
(with $\f = \lambda + \rho$) is
\be
\dim V_{g,k} = \sum_{\la\in\Lambda_{k}}\prod_{\a}\left(1-\ex{i\frac{\a(\la
+ \rho)}{k+h}}\right)^{1-g}\;\;,
\ee
again in agreement with the Verlinde formula.

\subsection*{Acknowledgements}

We thank the organizers of the School for providing us with the opportunity
to present these lectures.
We are also grateful to E. Witten for opening our eyes to the topological
intricacies behind our use of the Weyl integral formula.

\appendix

\section{Some Lie Algebra Theory}

This will be a lightning review of the basics of Lie algebra theory.
Recall that the Lie algebra $\lg$ may be decomposed into a maximally
commuting subalgebra $\lt$ (the Cartan subalgebra) and its complement
$\lk$ (which is a sum of non-trivial representations of $\lt$),
\be
\lg = \lt \oplus \lk\, .
\ee
The complexified Lie algebra $\lg_{\CC}$ of a compact group $G$
may be decomposed as
\be
\lg_{\CC} = \lt_{\CC} \oplus \oplus_{\a} \lg_{\a} \, ,
\ee
where the sum is over all roots $\a$ (not zero) and $\lt_{\CC}$ is the
complexified Cartan subalgebra. The root spaces $\lg_{\a}$ are the one
dimensional eigenspaces of the isotropy representation, meaning, for $t \in
\lt$ and $E_{\a} \in \lg_{\a}$
we have
\be
[t,E_{\a}] = \a(t) E_{\a} \, .
\ee
We have, correspondingly, a decomposition of $\lg$ as
\be
\lg = \lt \oplus \RR (E_{\a}+iE_{-\a}) \oplus \RR  (iE_{\a} - E_{-\a})
\, ,
\ee
where the roots are positive.

The coroots, $h_{\a}$, are defined by
\be
h_{\a} = [E_{\a},E_{-\a}] \, .
\ee
There is some latitude in the definition of the inner product on the
coroot space. For simple Lie algebras all are proportional to the
Killing form. Our choice for $SU(n)$, which is simply laced, is
\be
<h_{\a}, h_{\a}> =2 \, .
\ee
The resulting identification $\lt^{*} \sim \lt$ makes $h_{\a}$ correspond
with $\a$. The preferred inner product is
\be
<A,B> = - \Tr (AB) \, .
\ee
where the matrices are taken to be anti-Hermitian. For all the roots one has
\be
<\a_{i},\a_{j}> \, = \, (2\d_{ij}-\d_{i\, j+1}-\d_{i+1\, j}) \, .
\label{in1}
\ee

The positive roots are further decomposed into simple and not simple
roots. The simple roots are those positive roots that can not be written
as the sum of positive roots with integer positive co-efficients. The
number of simple roots is the same as the rank $r$ (dimension of $\lt$) of
the Lie algebra.

\noindent \underline{The Weyl Group}

We look at the action of the Weyl group on the roots. The Weyl group
acts by reflection through roots, so that for a weight $M$ the element
$S_{\a}$ of the Weyl group acts by
\be
S_{\a}(M)= M \,-\, 2\frac{<M,\a >}{<\a,\a >}\a \, .
\ee
The Weyl group itself is the group obtained on taking all combinations
of these reflections applied successively. It is enough to consider
those $S_{\a_{i}}$ where $\a_{i}$ are simple, as these will generate the
entire group.

The simple roots $\a_{i}$ determine fundamental weights $\la^{j}$ by
\be
\Tr \a_{i} \la^{j} = \d_{i}^{j} \, .
\ee

After these formalities it is time for
some examples.

\noindent \underline{$SU(2)$}

In this case we may take the positive root $\a$ and the fundamental
weight $\l$ to be
\be
\a = \left( \begin{array}{clcr}
 1 & 0 \\
 0 & -1
\end{array}
\right) \, , \; \;
\la =  \frac{1}{2}\left( \begin{array}{clcr}
 1 & 0 \\
 0 & -1
\end{array}
\right) \, .
\ee
The Weyl vector $\rho = \a /2 = \la$ and $tr \a ^{2}=2$. The Weyl group
has two elements $S_{\a}$ and the identity ($S_{\a}^{2}= 1$) and its action
is given by,
\be
S_{\a}(\a) = -\a.
\ee
Clearly the Weyl group is $\ZZ_{2}$.


\noindent \underline{$SU(3)$}

There are three positive roots for $SU(3)$, which we take to be
\be
\a_{1} = \left( \begin{array}{clcr}
 1 & 0 & 0\\
 0 & -1&0 \\
 0 & 0 &0
\end{array}
\right) , \; \;
\a_{2} = \left( \begin{array}{clcr}
 0 & 0 &0\\
 0 & 1&0 \\
 0 & 0 & -1
\end{array}
\right) , \; \;
\a_{3} = \left( \begin{array}{clcr}
 1 & 0 &0\\
 0 & 0&0 \\
 0 & 0 & -1
\end{array}
\right) \, .
\ee
With these definitions it is apparent that $\a_{3} =\a_{1} + \a_{2}$ so
that $\a_{1}$ and $\a_{2}$ are simple roots. The fundamental weights and
the Weyl vector are
\be
\la^{1} = \frac{1}{3}\left( \begin{array}{clcr}
 2 & 0 &0\\
 0 & -1&0 \\
 0 & 0 & -1
\end{array}
\right) , \; \;
\la^{2} = \frac{1}{3}\left( \begin{array}{clcr}
 1 & 0 &0\\
 0 & 1&0 \\
 0 & 0 & -2
\end{array}
\right) \; \;
\rho = \left( \begin{array}{clcr}
 1 & 0 &0\\
 0 & 0&0 \\
 0 & 0 & -1
\end{array}
\right)  \, .
\ee
The Weyl vector $\rho = (\a_{1}+\a_{2})/2 +\a_{3}/2 = \a_{3}$. The two
generators of the Weyl group are $S_{\a_{1}}$ and $S_{\a_{2}}$ and they
act by
\be
S_{\a_{i}}(\a_{j}) = - \a_{i}\d_{ij} + \a_{3}\d_{3 \, i+j} \, , \; \;
i,j = 1,2 \, .
\ee
The Weyl group in this case has $6$ elements and is the permutation group
$S_{3}$.

\section{Determinants}

We will deal with the determinants that arise in the Yang-Mills and
$G/G$ in a unified way. This is perhaps a little unnatural as in the
latter one must pick a complex structure on $\Sg$ while in the former
there is no real need to do this. In any event the answers do not depend
on this choice.

\subsection{The Dolbeault Complex}

The first thing to note is that at the points where we require the
ratios of determinants in the text, namely (\ref{detym}) and around
(\ref{g4}), the gauge fixing has only been partial. We had been careful
to preserve the Abelian $\bT$ invariance.
We should thus regularize in a manner which respects this residual
gauge invariance and we will accomplish this by using a heat kernel
(or $\zeta$-function) regularization based on the $\lt$ covariant
Laplacian $\Delta_{A} = -(d_{A}^{*}d_{A}+d_{A}d_{A}^{*})$ where
$A$ is the $\bT$ gauge field. For an operator $\cal O$ we set
\be
\log\Det{\cal O} = \Tr e^{-\eps\Delta_{A}}\log{\cal O}\;\;,
\ee
where we now use $\Tr$ to denote a functional trace (e.g.~including
an integration).

We begin with the determinants that arise on integrating out
$A^{\lk}$. For the $G/G$ theory this is given in (\ref{g4}) and up to an
overall factor, the relevant part of the action is
(using the differential form version (\ref{diff}))
\be
\tr( A^{\lk}*A^{\lk} - A^{\lk}t^{-1}(i+*)A^{\lk}\,t)\;\;.\label{trk}
\ee
On the other hand in the Yang-Mills theory this takes the simple form
\be
\tr A^{\lk}[\f^{\lt}, A^{\lk}] \, . \label{trkym}
\ee
To put these into more explicit forms we recall that, on the
root space $\lg_{\a}\ss\lk_{\CC}$, ${\rm Ad}(t)$ acts by multiplication
by $\exp i\a(\f)$. Furthermore, with respect to the Killing-Cartan
metric (trace), $\lg_{\a}$ and $\lg_{\beta}$ are orthogonal unless
$\beta = -\alpha$. Thus, expanding $A^{\lk}$ in terms of basis vectors
$E_{\a }$ of $\lg_{\a}$ such that
\be
A^{\lk}= \, i \sum_{\a}E_{\a}A^{\a}\;\;,\;\;\;\;\;\;\tr(E_{\a}E_{-\a}) = 1\;\;,
\ee
we can break up (\ref{trk}) into a
sum of terms depending only on the pair $\pm\a$. We obtain
\bea
(\ref{trk}) &=& \sum_{\a}A^{\a}*A^{-\a}-A^{\a}e^{-i\a(\f)}(i+*)A^{-\a}
\nonumber\\
&=& \sum_{\a>0}\left[ A^{\a}(i+*)M_{-\a}A^{-\a}
           - A^{\a}(i-*)M_{\a}A^{-\a}\right]\;\;, \label{split}
\eea
where $M_{\a}$ is the number
\bea
M_{\a}&=&\left(1-e^{i\a(\f)}\right)\;\;,\nonumber\\
\prod_{\a}M_{\a}&=&\det ({\bf 1}-{\rm Ad}_{\lk}(e^{i\f}))\;\;.\label{s9}
\eea
We can express (\ref{trkym}) in exactly the same way except now in
(\ref{split}) the numbers $M_{\a}$ are
\bea
M_{\a}&=& i\a(\f) \, , \nonumber \\
\prod_{\a} M_{\a} &=& \det{}_{\lk}{ (\ad(\f))} \, .
\eea

Writing these in terms of the scalar product on $1$-forms,
one sees that the path integral over $A^{\lk}$ yields
\be
\prod_{\a>0}\Det\left[(1+i*)M_{\a} + (1-i*)M_{-\a}\right]^{-1}\;\;.
\label{s1}
\ee
Here we recognize the projectors
\be
P_{\pm} = \trac{1}{2}(1\pm i*)
\ee
onto the spaces of $(1,0)$-forms ($\sim dz$) and $(0,1)$-forms
($\sim d\bar{z}$) respectively and thus (\ref{s1}) exhibits
quite clearly the chiral nature of the (gauged) WZW model. For the
Yang-Mills theory, on the other hand, the $*$ components of (\ref{s1})
cancel and the determinant is certainly diagonal.

As a
consequence of the presence of the projectors $P_{\pm}$ in (\ref{s1}),
the two summands act on different spaces. For each $\a$ we may thus
write the determinant as a product of the $(1,0)$ and $(0,1)$
pieces,
\be
\Det\left[(1+i*)M_{\a} + (1-i*)M_{-\a}\right]^{-1}
=\left[\Det_{(1,0)}M_{\a}\right]^{-1} \times
\left[\Det_{(0,1)}M_{-\a}\right]^{-1}\;\;.\label{s2}
\ee
Before evaluating this, we will combine it with the contributions
from the ghosts (equivalently, the Weyl integral formula). The ghost
action has the form
\be
\sum_{\a>0}\left[\bar{c}^{\a}*M_{-\a}c^{-\a}
    + \bar{c}^{-\a}*M_{\a}c^{\a}\right]     \label{s3}\;\;,
\ee
and therefore the ghost determinant is
\be
\prod_{\a>0}\Det_{0}M_{\a}\Det_{0}M_{-\a}\label{s4}\;\;.
\ee
Combining this with (\ref{s2}), we see that we need to determine
and make sense of
\be
\prod_{\a>0}\left[\Det_{0}M_{\a}\Det^{-1}_{(1,0)}M_{\a}\right]
          \left[\Det_{0}M_{-\a}\Det^{-1}_{(0,1)}M_{-\a}\right]\;\;.
\label{s5}
\ee
This we will accomplish by relating the products of these determinants
to the Witten index of the Dolbeault complex. Indeed, suppose that $M_{\a}$
is a constant. Then
\be
\log\Det_{0} M_{\a}\Det^{-1}_{(1,0)}M_{\a} =
\left[\Tr_{0}e^{-\eps\Delta_{A}} - \Tr_{(1,0)}e^{-\eps\Delta_{A}}\right]
\log M_{\a}\;\;,\label{s6}
\ee
where we need to remember that the Laplacian $\Delta_{A}$ acts to the
right on one-forms taking values in $\lg_{(-\a)}$,
the root space of $(-\a)$. There we have
\be
d_{A}|_{(-\a)} = d -i\a (A) \equiv d + \tr(\a\a_{l})A^{l}\;\;,
\ee
so that the `charge' is $\tr(\a\a_{l})$.
The term in brackets is nothing but the index of the Dolbeault complex,
\be
\left[\Tr_{0}e^{-\eps\Delta_{A}} - \Tr_{(1,0)}e^{-\eps\Delta_{A}}\right]
=\sum_{p=0}^{1}(-1)^{p}b^{p,0} = {\rm Index}\;\bar{\del}_{A}\;\;.
\ee
This index can of course be calculated directly from the heat kernel
expansion, but one may as well
call upon the known result that for the Dolbeault operator coupled to
a vector bundle $V$ with connection $A$ one has (see e.g.~\cite{gi})
\be
{\rm Index}\;\bar{\del}_{A} = \int_{M}{\rm Td}(T^{(1,0)}(M)){\rm ch}(V)\;\;.
\ee
In two dimensions this reduces to
\be
{\rm Index}\;\bar{\del}_{A}=\trac{1}{2}\c(\S) + c_{1}(V)\;\;. \label{dolind}
\ee
Therefore, in the case at hand, one finds that (\ref{s6}) equals
\bea
{\rm Index}\;\bar{\del}|_{(-\a)} \log M_{\a} &=& \left[\trac{1}{2}\c(\S) +
c_{1}(V_{(-\a)})\right]\log M_{\a}\nonumber\\
&=&\left[\trac{1}{8\pi}\int_{\S}R + \trac{1}{2\pi}\int_{\S}\tr(\a\a_{l})F^{l}
\right]\log M_{\a}\;\;.
\eea

When $M_{\a}$ is not a constant, one simply has to move $\log M_{\a}$
into the integral, so that one obtains
\be
\log\Det_{0} M_{\a}\Det^{-1}_{(1,0)}M_{\a} =
\trac{1}{8\pi}\int_{\S}R\log M_{\a} +
 \trac{1}{2\pi}\int_{\S}\tr(\a\a_{l})F^{l}\log M_{\a}\;\;.\label{s7}
\ee
To see that this is correct, we write
\bea
\Tr\log M_{\a} e^{-\eps\Delta_{A}} &\equiv& \int\!dx\,
\langle x|\log M_{\a} e^{-\eps\Delta_{A}}|x\rangle\nonumber\\
&=&\int\!dx\,\log M_{\a}(x)\langle x|e^{-\eps\Delta_{A}}|x\rangle\;\;,
\eea
and note that $R$ and $F$ arise as the first Seeley coefficients in the
expansion of $\langle x|e^{-\eps\Delta_{A}}|x\rangle$. It is worthwhile
remarking that
the result (\ref{s7}) is {\em finite}, the $1/\eps$ poles cancelling
between the scalar and one-form contributions.

One can proceed analogously for the second factor in (\ref{s5}). In this
case it is the index of $\del_{A}$ that makes an appearance and which
differs by the sign of the second summand from (\ref{dolind}),
\be
{\rm Index}\;\del_{A} = \trac{1}{2}\c(\S) - c_{1}(V)\;\;.
\ee
As the Laplacian still acts on $\lg_{(-\a)}$, one obtains a
$\log M_{\a} - \log M_{-\a}$ contribution to the gauge field part of the
index, while it is the sum of the two terms that contributes to
the gravitational part. Hence one finds that the regularized determinant
(\ref{s5}) is
\be
(\ref{s5}) = \prod_{\a>0}
\exp\left(\trac{1}{8\pi}\int_{\S}R\log M_{a}M_{-\a}
+ \trac{1}{2\pi}\int_{\S}\tr(\a\a_{l})F^{l}\log\frac{M_{\a}}{M_{-\a}}
\right)\;\;.
\label{s8}
\ee
We will now consider separately the two contributions to this expression.

\subsection{The Weyl Determinant}

\noindent
Rewriting the term in (\ref{s8}) that depends on the curvature as
\be
\exp\left[\trac{1}{8\pi}\int_{\S}R\sum_{\a>0} \log M_{\a} M_{-\a}\right]\;\;,
\label{s10}
\ee
one recognizes it as a dilaton like coupling to the metric, the role of the
dilaton being played by
\bea
\sum_{\a>0}\log M_{\a} M_{-\a}& =&   \log\det({\bf 1} - {\rm
Ad}(e^{i\f})) \; \; {\rm for\; the \; G/G \; model} \nonumber \\
&=& \log\det{}_{\lk}(\ad(\f)) \; \; \; \; \; \; \; {\rm for \; Yang-Mills
\; theory.}
\eea
If $\f$ is constant (\ref{s10}) reduces to
\be
\left[\det({\bf 1} - {\rm Ad}(e^{i\f}))\right]^{\c(\Sg)/2} \; \; {\rm or}
\; \; \; \; \left[\det{}_{\lk}({\rm ad}(\f))\right]^{\c(\Sg)/2}\, .
\ee

\subsection{The Shift $k\ra k+h$}

\noindent
We now come to the crux of the matter. The second term in (\ref{s8})
is responsible for the shift in $k$ in the $\bG /\bG$ theories. To see
this note that
\be
\frac{M_{\a}}{M_{-\a}} = \frac{1-e^{i\a\f}}{1-e^{-i\a\f}} = - e^{i\a\f}
\;\; , \label{ratio}
\ee
so that
\be
\prod_{\a>0}\exp\left(\trac{1}{2\pi}\int_{\S}\tr(\a\a_{l})F^{l}\log
  \frac{M_{\a}}{M_{-\a}}\right) = \exp\left[\trac{i}{2\pi}\sum_{\a>0}
\int_{\S}\tr(\a\a_{l})\a(\f)F^{l}\right]\;\;. \label{s11}
\ee
Here we have suppressed the imaginary contribution to the $\log$, as it
will make no appearance for simply connected groups (where $\rho =
\frac{1}{2}\sum_{\a >0}\a$ is integral).

We now put the exponent in more manageable form by noting that the
(negative of the) Killing-Cartan metric $b$ of $\lg$, restricted to $\lt$,
\be
b(X,Y) = - \tr {\rm ad}(X)\,{\rm ad}(Y)\;\;,
\ee
can be written in terms of the roots as
\be
b(X,Y) = 2\sum_{\a>0}\a(X)\a(Y)\;\;.
\ee
Moreover, with our convention that ${\rm ad}(X)|_{\a} = i\a(X)$,
$b(X,Y)$ is related to the Coxeter number (or quadratic Casimir of the
adjoint representation) $h$   via
\be
b(X,Y)= 2h \tr(XY)
\ee
($h=n$ for $SU(n)$). Hence the exponent becomes
\bea
\trac{i}{2\pi}\sum_{\a>0}\int_{\S}\tr(\a\a_{l})\a(\f)F^{l} &=&
\trac{i}{4\pi}\int_{\S} b(\f,\a_{l})F^{l}\nonumber\\
&=& \trac{ih}{2\pi}\int_{\S}\f_{l}F^{l}\;\;,
\eea
which produces precisely the long awaited
shift $k\ra k+h$ in the action $S_{\f F}$,
\be
\trac{ik}{2\pi}\int_{\S}\f_{l}F^{l} \longrightarrow
\trac{i(k+h)}{2\pi}\int_{\S}\f_{l}F^{l}\;\;.\label{shift}
\ee

On the other hand, in the case of Yang-Mills theory no such shift
occurs. The equivalent of (\ref{ratio}) is
\be
\frac{M_{\a}}{M_{-\a}} = \frac{i\a(\f)}{-\a(\f)} = -1 \, ,
\ee
which gives a vanishing contribution for simply connected groups. This
is as it should be. We could have used the De Rham complex in this case
and twisting with a vector bundle does not change the index (except to
multiply it by the dimension of the vector space) which just
depends on the Euler character of the manifold.


\rnc{\Large}{\normalsize}

\end{document}